\documentclass{aa}
\usepackage{multirow}
\usepackage{amsmath}
\usepackage[colorlinks,linkcolor=blue,anchorcolor=blue,citecolor=blue]{hyperref}
\usepackage{graphicx} 
\usepackage{lscape}
\usepackage{longtable}
\usepackage{txfonts}
\usepackage{natbib}
\usepackage{color}

\begin{document}

\title{The SRG/eROSITA All-Sky Survey: Exploring halo assembly bias with X-ray selected superclusters} 

\author{
A.~Liu\inst{1}\thanks{e-mail: \href{mailto:liuang@mpe.mpg.de}{\tt liuang@mpe.mpg.de}},
E.~Bulbul\inst{1},
T.~Shin\inst{2},
A.~von~der~Linden\inst{2,1},
V.~Ghirardini\inst{1},
M.~Kluge\inst{1},
J.~S.~Sanders\inst{1},
S.~Grandis\inst{3},
X.~Zhang\inst{1},
E.~Artis\inst{1},
Y.~E.~Bahar\inst{1},
F.~Balzer\inst{1},
N.~Clerc\inst{4},
N.~Malavasi\inst{1},
A.~Merloni\inst{1},
K.~Nandra\inst{1},
M.~E.~Ramos-Ceja\inst{1},
S.~Zelmer\inst{1}
}
\institute{
\inst{1}{Max Planck Institute for Extraterrestrial Physics, Giessenbachstrasse 1, 85748 Garching, Germany}\\
\inst{2}{Department of Physics and Astronomy, Stony Brook University, Stony Brook, NY 11794, USA}\\
\inst{3}{Institute for Astro- and Particle Physics, University of Innsbruck, Technikerstr. 25, 6020 Innsbruck, Austria}\\
\inst{4}{IRAP, Universitt’e de Toulouse, CNRS, UPS, CNES, F-31028 Toulouse, France}
}

\titlerunning{Exploring HAB with eRASS1 superclusters}
\authorrunning{Liu et al.}

\abstract
{Numerical simulations have indicated that the clustering of dark matter halos is not only dependent on their masses but has a secondary dependence on other properties, such as the assembly history of the halo. This phenomenon, known as the ``halo assembly bias (HAB)'', has been found mostly on galaxy scales, lacking observational evidence on larger scales. In this work, we propose a novel method to explore HAB on cluster scales using large samples of superclusters. Leveraging the largest-ever X-ray galaxy cluster and supercluster samples obtained from the first {\sl SRG/eROSITA} all-sky survey, we construct two subsamples of galaxy clusters which consist of supercluster members (SC) and isolated clusters (ISO) respectively. After correcting the selection effects on redshift, mass, and survey depth, we compute the excess in the concentration of the intracluster gas of isolated clusters with respect to supercluster members, defined as $\delta c_{\rm gas} \equiv c_{\rm gas,ISO}/c_{\rm gas,SC}-1$, to investigate the environmental effect on the concentration of clusters, an inference of halo assembly bias on cluster scales. 
We find that the average gas mass concentration of isolated clusters is a few percent higher than that of supercluster members, with a maximum significance of $2.8\sigma$. The result on $\delta c_{\rm gas}$ varies with the choices on the overdensity ratio $f$ in supercluster identification, cluster mass proxies, and mass and redshift ranges, but remains positive in almost all the measurements. We measure slightly larger $\delta c_{\rm gas}$ when adopting a higher $f$ in supercluster identification. $\delta c_{\rm gas}$ is also larger for low-mass and low-redshift clusters. We perform weak lensing analyses to compare the total mass concentration of the two classes and find a similar trend in total mass concentration as obtained from gas mass concentration. Our results are consistent with the prediction of HAB on cluster scales, where halos located in denser environments are less concentrated, and this trend is stronger for halos with lower mass and at lower redshifts. These phenomena can be interpreted by the fact that clusters in denser environments such as superclusters have experienced more mergers than isolated clusters in their assembling history. This work paves the way to explore HAB with X-ray superclusters and demonstrates that large samples of superclusters with X-ray and weak-lensing data can advance our understanding of the evolution of large-scale structures.  }


\maketitle

\section{Introduction}
\label{sec:intro}
In the standard $\Lambda$CDM structure formation picture, galaxies and galaxy clusters form and evolve following their host dark matter halos, which emerge from local density peaks of initial fluctuations. The spatial distribution of galaxies and clusters is therefore a powerful tool to test the structure formation scenario. According to the excursion set theory, the clustering of dark matter halos is biased by the halo mass: more massive halos have higher clustering magnitudes than the overall clustering of mass \citep[e.g.,][]{1974Press,1991Bond,1996Mo,2007Zentner}. On the other hand, since the beginning of the 21st century, large $\Lambda$CDM $N-$body simulations have indicated that the clustering of dark matter halos is not only dependent on their masses but has a secondary dependence on other properties, such as halo formation time, concentration, shape, and spin \citep[e.g.,][]{2004Sheth,2005Gao,2006Wechsler,2007Jing,2008Li}. This phenomenon, commonly referred to as ``halo assembly bias (HAB)'' \citep{2007Gao}, challenges the standard halo occupation modeling, which assumes that the halo occupation distribution (HOD), defined as the probability a halo hosts a given number of galaxies, depends only upon halo mass. Although the effect of HAB is small compared to the mass-dependent halo bias, it can significantly bias the HOD, and induce a non-negligible systematic error in the study of galaxy evolution and precision cosmology \citep[e.g.,][]{2014Zentner}.  

Numerous works have been performed to study HAB using both simulations and galaxy surveys. Thanks to these efforts in the past 20 years, a general picture has been established to describe HAB in low-mass halos, namely, galaxy-sized halos. In this picture, the clustering magnitude of halos is correlated with halo formation time, where old halos are more clustered than younger ones \citep[see, e.g.,][among many others]{2005Gao,2006Zhu,2006Yang,2006Harker,2007Wetzel,2007Gao,2007Jing,2009Fakhouri,2012vanDaalen}. This trend is commonly found in many works, despite that the proxy for halo formation time may differ from one to another. Concentration is often used to quantify the halo formation time, with older halos being more concentrated \citep[e.g.,][]{2006Wechsler}. Other proxies such as stellar age \citep[e.g.,][]{2014Lacerna}, stellar-mass assembly history \citep[e.g.,][]{2017Montero-Dorta}, specific star formation rate \citep{2013Wang}, and the redshift at which the halo has assembled a certain fraction of its final mass \citep[see][]{2007Jing,2008Li}, generally show the same trend.

The theoretical origins of HAB have been discussed since its first detection. Although HAB is often described as the dependence of clustering magnitudes on halo properties, it is more intuitive to attribute this ``dependence'', or more accurately, ``correlation'', to the impact of the large-scale environment of a halo on its properties. A commonly accepted scenario for the HAB of low-mass halos is that the bias is likely due to the tidal fields of the neighboring massive halos, which suppress the mass accretions of the low-mass halos \citep[e.g.,][]{2007Wang}. This scenario is supported by many works from both simulations and observations \citep{2007Diemand,2009Hahn,2009Ludlow,2016Fang,2017Borzyszkowski,2018Paranjape,2018Salcedo,2019Ramakrishnan,2021Rodriguez}. Besides that, some other works also suggest that the detection or non-detection of HAB is dependent on the definition of halo mass \citep[e.g.,]{2017Villarreal,2018Chue}. 

Compared to the results in low-mass regimes, the situation for high-mass halos, namely, cluster-sized halos, is less clear. Several works indicate that the scaling between halo clustering magnitude and concentration changes its sign around the characteristic collapsing mass $M_{\ast}$\footnote{$M_{\ast}$ is defined as the mass where $\sigma(M_{\ast})\approx 1.686$. For Millennium Simulations, $M_{\ast}$ is around $10^{13}M_{\odot}$ }. More concentrated halos are more strongly clustered for halos below $M_{\ast}$, as mentioned above, which is reversed for halos above $M_{\ast}$ \citep[see, e.g.,][]{2006Wechsler,2007Wetzel,2007Jing,2007Gao,2010Faltenbacher}. This contradicts several other works, where HAB is found to be weak or even absent in groups, clusters, or massive galaxies \citep[e.g.,][]{2008Wang,2008Li,2016Lin,2017Dvornik,2018Mao,2019Zentner}. In fact, if large-scale tidal fields are the main physical origin of HAB, then we should not expect strong HAB for massive halos \citep[e.g.,][]{2005Mo}, as they are much less affected by these tidal fields \citep[however, see][in which the statistics of the peaks of Gaussian random fluctuations are proposed as an origin of HAB in massive halos]{2008Dalal}.
A strong detection of HAB in clusters is reported in \citet{2016Miyatake} and \citet{2016More}, using the projected separation of potential member galaxies as the proxy for halo age. However, these results remain controversial as the detected HAB signal is probably due to the projection effect of line-of-sight large-scale structures \citep[e.g.,][]{2017Zu,2019Sunayama}. More recently, \citet{2022Lin} reported a $3\sigma$ detection of HAB using the simulated counterpart halos of 634 massive clusters identified with SDSS. In summary, the detection of HAB in cluster-sized halos is still inconclusive, either from simulations or observations.

There are a few challenges in detecting HAB in observations. First, one needs to find an observable as a reliable proxy for halo assembly history. These include the aforementioned concentration, stellar age, stellar assembly history, etc. 
Another challenge is to establish a link between this observable and halo clustering magnitude or large-scale environment, after accounting for all the systematic biases. 

\begin{figure*}
\begin{center}
\includegraphics[height=0.59\textwidth, trim=0 0 -20 0, clip]{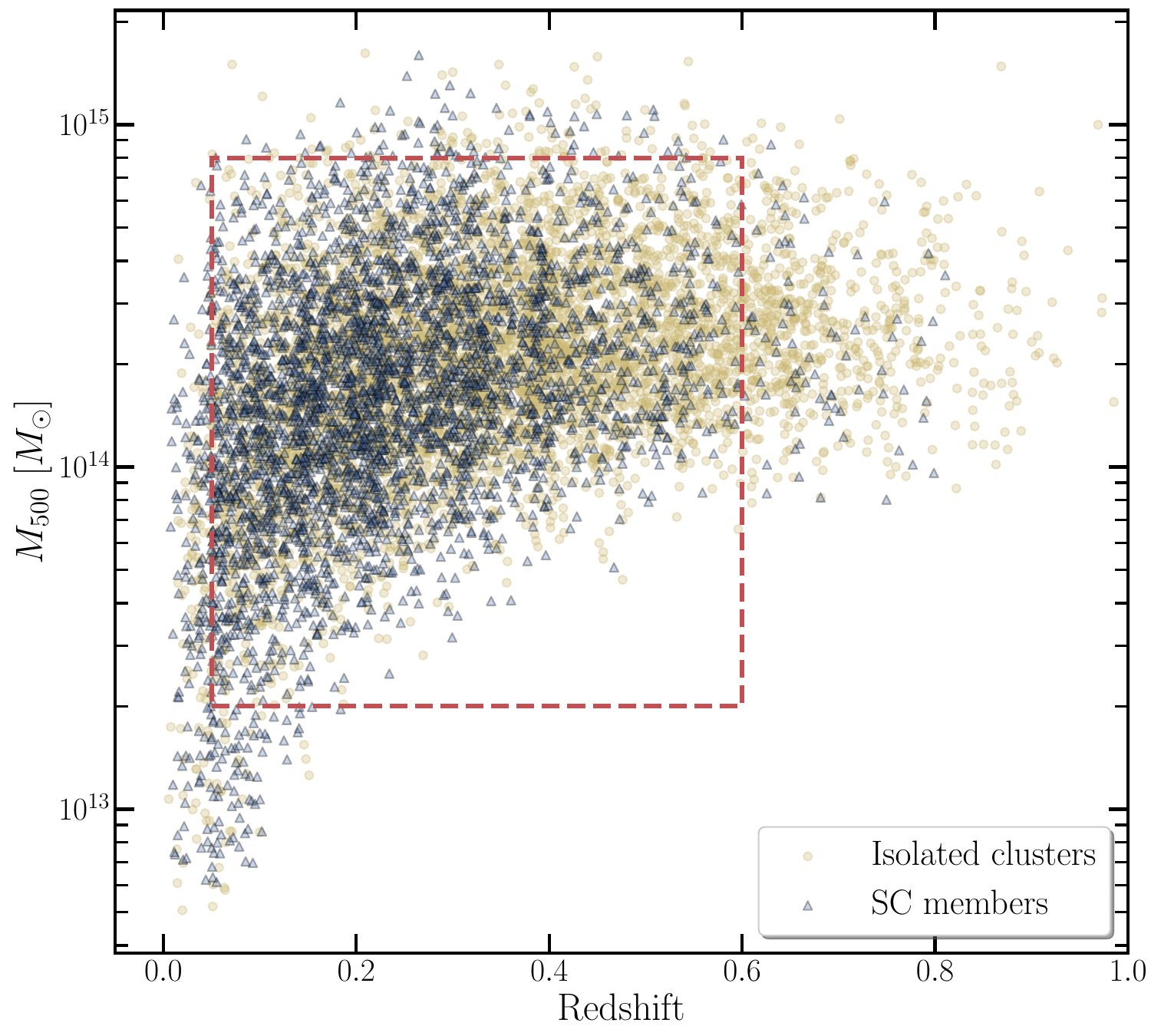}
\includegraphics[height=0.59\textwidth, trim=0 0 10 0, clip]{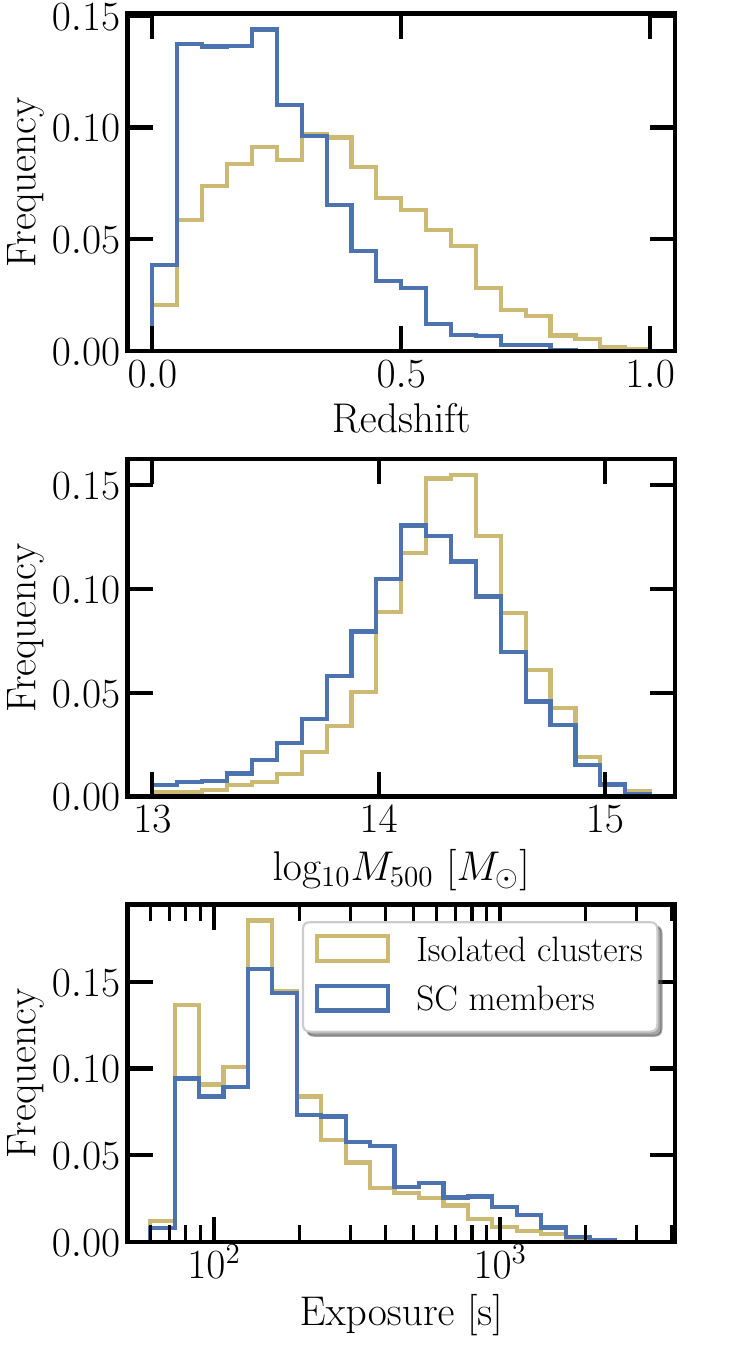}
\caption{\label{fig:sample} Comparison of the properties of the two subsamples before trimming: blue for supercluster members and yellow for isolated clusters. The red rectangle in the left panel defines the $M-z$ space where we select the subsamples for further selection (trimming). The histograms in the right panel show the distribution of redshift, mass, and exposure time of the two subsamples before the trimming process. }
\end{center}
\end{figure*}

\begin{figure*}
\begin{center}
\includegraphics[width=0.99\textwidth, trim=0 30 5 70, clip]{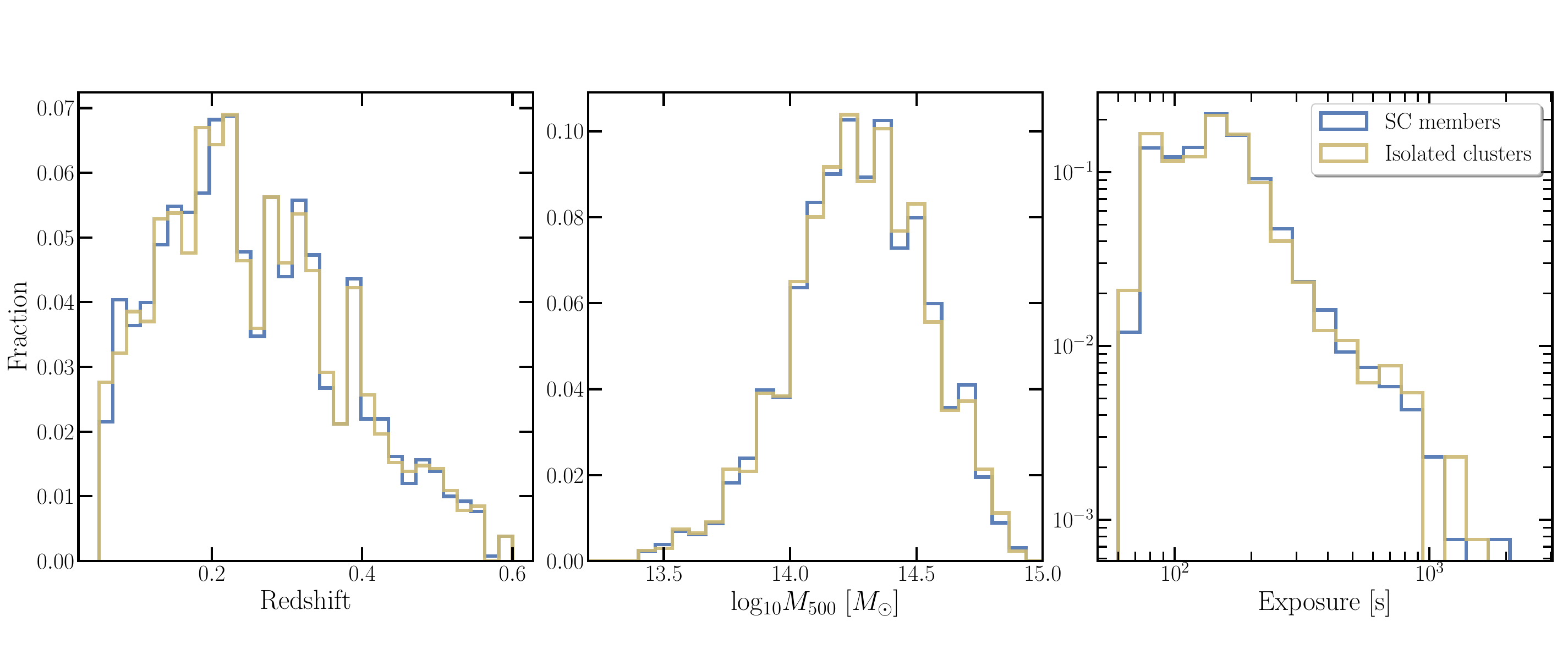}
\caption{\label{fig:mzt_compare} Comparison of the properties of the two subsamples after trimming. The histograms show the distribution of redshift, mass, and exposure time, from left to right. Supercluster members and isolated clusters are plotted in blue and yellow, respectively. }
\end{center}
\end{figure*}

In most of the previous works, the strategy to detect HAB in cluster-sized halos is to split the clusters into subsamples according to the assembly history and compare the clustering magnitudes in these subsamples. Alternatively, one could also adopt another strategy, that is to (1) divide the clusters into different classes based on their clustering magnitudes or environments and (2) compare the assembly history (or, in practice, the observational proxies of assembly history) of these classes of clusters. For the first step, superclusters (SCs) can be used as an indicator of the clustering magnitudes or environments of galaxy clusters: by definition, the clusters that are members of supercluster systems have higher clustering magnitudes than the ones that are not (namely, isolated clusters); Supercluster members are also located in a denser environment than isolated clusters. Provided that the two classes of clusters are consistent in other essential sample properties, such as redshift, mass, and selection function, the difference in assembly history proxies between them should be regarded as direct evidence for HAB. 

Apart from the previously mentioned average member galaxy separation, a few other quantities such as concentration can also be used as observable proxies for the halo assembly history of clusters \citep[e.g.,][]{1997Navarro,Wechsler2002,2006Lu,2021Lau}. While the concentration of the total mass is measurable through weak lensing analysis using galaxy survey data, the concentration of the intracluster medium (ICM) can be measured in the X-ray band thanks to the bremsstrahlung emission from the ICM. Particularly, ICM properties measured from X-rays have the advantage over the ones from optical galaxies, because ICM is a single and continuous object in the X-ray band, whose distribution can be described accurately by parametric models, and is free of projection effects. The only caveat in using gas mass concentration as the proxy of halo assembly history is that nonthermal processes may reshape the distribution of gas in cluster core. For example, mechanical feedback activities of the central AGN often create cavities in the core of clusters and push the ICM from cluster center to larger radii \citep[see][for a review]{fabian2012}. However, if we define a reasonably large core radius, then these processes are not strong enough to significantly reduce the enclosed gas mass within the core. This can be inferred from the spatial distribution of metals within the ICM. The distribution of metals in the ICM can be described as a combination of a central peak and a large-scale plateau, where the central peak forms and evolves through the feedback activities of the central AGN \citep[e.g.,][]{Liu2019,liu2020}. Thus the extent of the central peak of metals in the ICM broadly traces the scope of action of AGN feedback. According to the numerous studies of ICM metallicity, in most clusters the size of the central metal peak is smaller than 0.3$R_{500}$\citep[see, e.g.,][]{liu2018,mernier2018,2021Gastaldello}, from which we can infer that the amount of ICM being moved from the center to $>0.3R_{500}$ due to AGN feedback is almost negligible. 

Therefore, the strategy of exploring HAB using superclusters can be summarized as comparing supercluster members and isolated clusters in the concentration of gas mass (from X-ray analysis) and total mass (from weak lensing analysis). 
To perform the above analysis, it is essential to have large samples of supercluster members and the corresponding isolated clusters\footnote{It should be noted that superclusters identified with X-ray cluster surveys are usually open structures without strict physical boundaries. In the most extreme case, the Universe can be regarded as a large ``supercluster''. Therefore, there is no absolute dividing line between ``supercluster members'' and ``isolated clusters''. The division of these two classes depends on the parent X-ray cluster sample and the criteria for supercluster identification. There are attempts to define superclusters as the systems that will survive the cosmic expansion and eventually collapse, based on their overdensities \citep[see, e.g.,][]{2015Chon}. However, predicting whether a supercluster will collapse from the estimation of its overdensity is highly uncertain. Even for the simplest cases of superclusters, namely, cluster pairs (except for those compact pairs in the merging process), observationally verifying such predictions would at least require accurate measurements of the peculiar velocities of the member clusters and their distance, which are not available for most of the currently known supercluster systems. Definitions for supercluster boundaries are also proposed in other domains, for example, optical galaxy surveys or simulations \citep[e.g.,][]{2019Einasto,2019Dupuy}. However, these definitions may not apply to X-ray cluster surveys. }, with X-ray data to measure the ICM properties and galaxy survey data to measure weak lensing shear profiles. Although superclusters have been known and studied for many years in both optical \citep[e.g.,][]{1989Scaramella,1993Zucca,1994Einasto,2007Einasto,2012Liivamagi,2024Chen} and X-rays \citep[e.g.,][]{2001Einasto,2013Chon,2014Chon,2018Adami,Boehringer2021,2022Liu}, such large samples and datasets are available only recently, thanks to the X-ray all-sky surveys conducted by the {\sl SRG/eROSITA} \citep[extended ROentgen Survey with an Imaging Telescope Array,][]{2021Predehl} X-ray telescope. In the first {\sl eROSITA} All-Sky Survey \citep[eRASS1,][]{Merloni2023}, 12247 galaxy clusters are detected and optically confirmed in the western Galactic hemisphere\footnote{Defined as ($179.9442^\circ < l < 359.9442^\circ$)}, up to redshift 1.32, with an estimated purity of 86\% \citep{Bulbul2024,Kluge2024}. Masses within $R_{500}$\footnote{$R_{500}$ is the radius within which the average matter density is 500 times the critical density at cluster’s redshift} of the clusters are computed based on the scaling relation between X-ray count rate, redshift, and mass, after being calibrated with weak lensing shear signal \citep[see][]{Bulbul2024,Grandis2024,Kleinebreil2024,Ghirardini2024}. 

In \citet{Liu2024}, we search for supercluster systems in eRASS1. We select a subsample of 8862 clusters from the eRASS1 cluster catalog with higher purity (96.4\%) and accurate redshifts ($\delta z/(1+z)<0.02$). A Friends-of-Friends (FoF) method is employed to identify supercluster systems that have 10 times higher cluster number density than the average density at the same redshift and survey depth. Using the above sample and method, we identify 1338 supercluster systems up to redshift 0.8, including 818 cluster pairs and 520 rich superclusters with $>2$ members. These supercluster systems enable us to split the 8862 selected eRASS1 clusters into two subsamples, which consist of 3948 supercluster members and 4914 isolated clusters, respectively. This large sample enables us to make a systematic comparison of the environmental effects on the properties of cluster-sized halos. 

In this work, we will explore HAB in cluster scales utilizing the eRASS1 galaxy cluster sample and supercluster sample, following the strategy described above. As the first attempt to study HAB with superclusters, the main objective of this work is to verify the feasibility of this direction, establish an effective methodology, and investigate possible caveats. The paper is organized as follows. In Sect.~\ref{sec:cluster}, we introduce the galaxy cluster sample and supercluster sample we used in this work. In Sect.~\ref{sec:results}, we compare the ICM mass concentrations of supercluster members and isolated clusters. In Sect.~\ref{sec:wl}, we compare the total mass concentrations of the two samples by performing weak lensing analysis. In Sect.~\ref{sec:discussions}, we interpret our results and discuss several caveats that may affect our analysis. Our conclusions are summarized in Sect.~\ref{sec:conclusions}. Throughout this paper, we adopt the concordance $\Lambda$CDM cosmology with $\Omega_{\Lambda} =0.7$, $\Omega_{\mathrm m} =0.3$, and $H_0 = 70$~km~s$^{-1}$~Mpc$^{-1}$. However, we note that the exact choice of cosmological parameters does not affect the results significantly. Quoted error bars correspond to a 1$\sigma$ confidence level.

\begin{figure*}
\begin{center}
\includegraphics[width=0.49\textwidth, trim=5 15 65 65, clip]{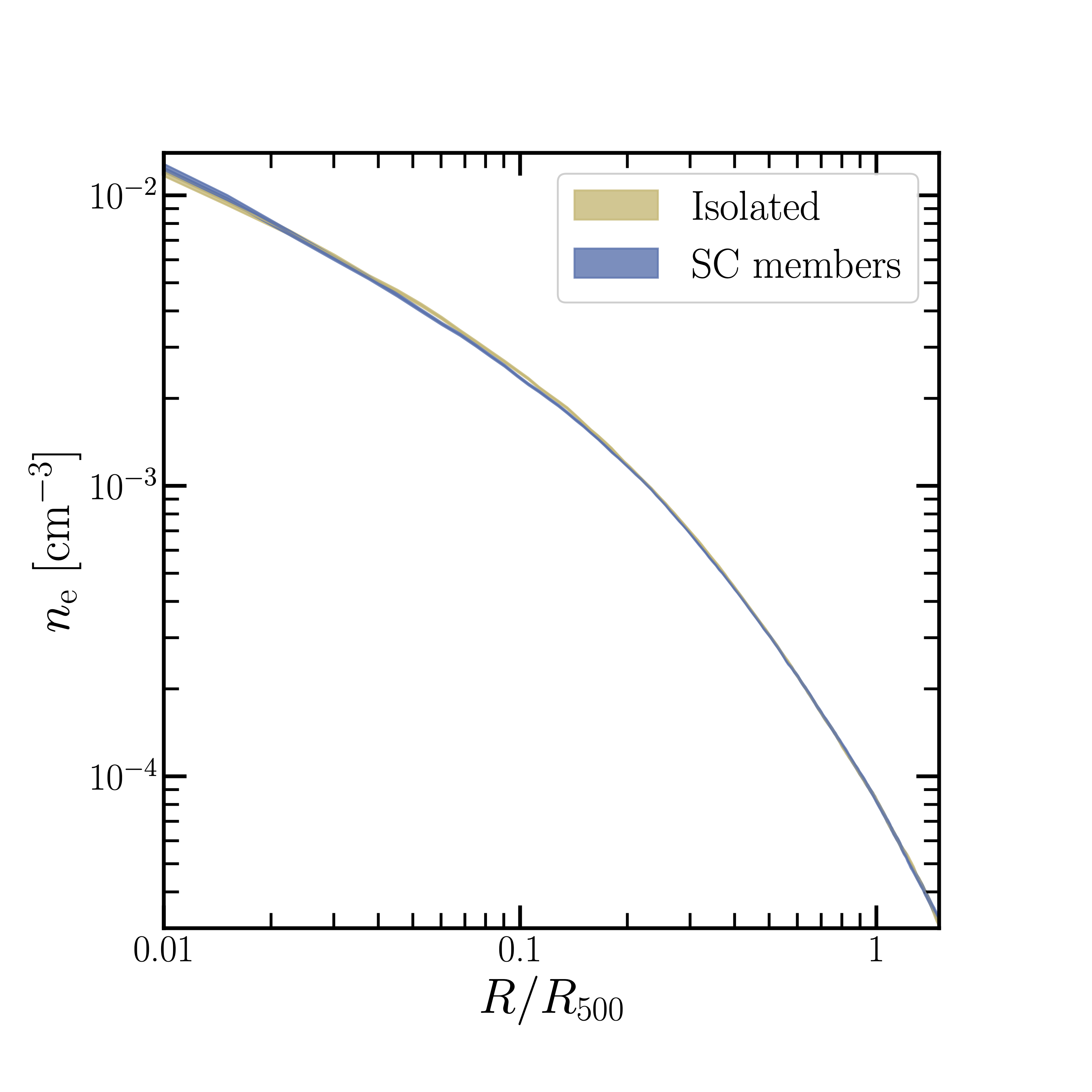}
\includegraphics[width=0.49\textwidth, trim=5 15 65 65, clip]{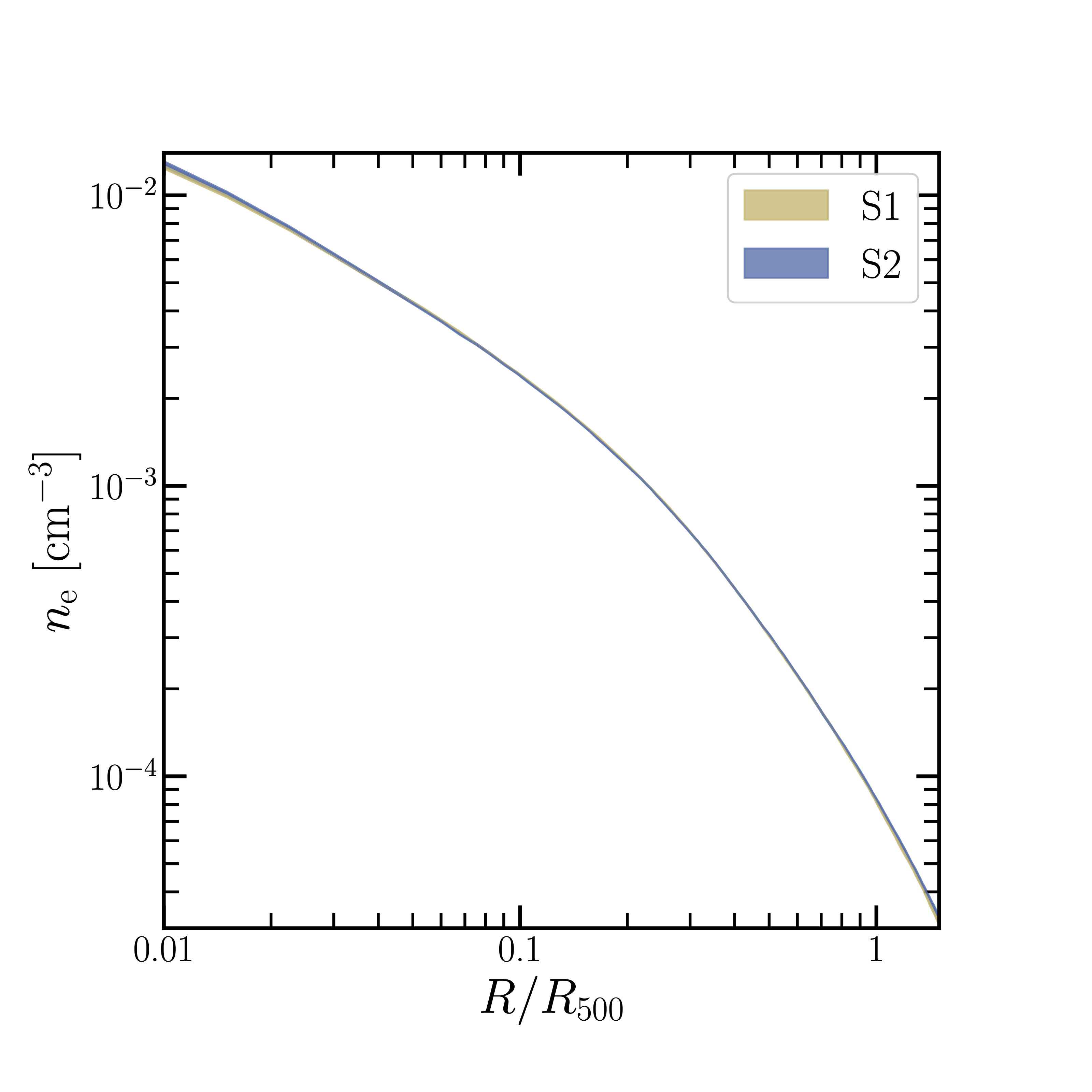}
\caption{\label{neprof} Median electron density profiles. In the left panel, the results for supercluster members and isolated clusters are plotted in blue and yellow, respectively. The uncertainties of the two profiles are almost negligible due to the large number of clusters in each sample. The difference in gas mass concentration between the two samples is almost invisible. In the right panel, we plot the results for the two control samples, which are selected by randomly dividing the cluster sample into two.  }
\end{center}
\end{figure*}

\section{Construction of the galaxy cluster samples}
\label{sec:cluster}

The eRASS1 supercluster sample naturally splits the parent cluster sample into two classes with significantly different clustering magnitudes and environments: the supercluster members are more clustered than isolated clusters and are located in denser environments. However, caution should be taken before we compare the properties of the two subsamples because the results can be biased by selection effects. To learn which selection effects are relevant to our analysis, we need to recall how the two subsamples are established, namely, the supercluster identification processes in \citet{Liu2024}. The linking length in the FoF algorithm is computed in a way such that the cluster density in a supercluster is $f$ times higher than the local density. The local density of eRASS1 detected clusters is dependent on redshift and survey depth. To correct this effect, we compute the cluster density (and thus also the linking length) as a function of both redshift $z$ and exposure time $t$:
\begin{equation}
\label{eqlk}
       l(z,t) = \left(\frac{N(z,t)}{V(z,t,A(t))}\cdot f\right)^{-1/3}.
\end{equation}
In Eq.~\ref{eqlk}, $N(z,t)$ denotes the number of clusters in the survey volume $V(z,t,A(t))$, where $A(t)$ is the eRASS1 survey area corresponding to the exposure time $t$. $f$ is the overdensity ratio.
This adaptive linking length has compensated for the deficit of superclusters at high redshifts and shallow survey areas. However, we unavoidably identify more supercluster members at lower redshifts and deeper areas. The redshift bias further induces a bias in mass, which should be corrected in the investigation of assembly bias. These selection effects can be observed from the right panels of Fig.~\ref{fig:sample}, where we compare the two classes of clusters in the distribution of redshift, $M_{500}$, and exposure time. 

In addition to the selection effects on redshift and mass, the fact that the clusters in the two subsamples are located at different survey depths also has non-negligible effects on the detection of assembly bias when concentration is adopted as the proxy for assembly history. A feature of X-ray surveys is that the selection of clusters is not a simple function of flux, but is affected by surface brightness patterns, such as the existence of a bright cool core, or more generally, the concentration or morphology of X-ray emission. The shallower the survey is, the more impact it suffers from these surface brightness patterns. This is because clusters with relatively flat surface brightness profiles have larger probabilities of being hidden in the background. As a comparison, cluster surveys based on Sunyaev–Zeldovich (SZ) effect are less sensitive to surface brightness patterns because the SZ signal is almost a linear function of $n_{\rm e}$, instead of $n_{\rm e}^2$ for X-ray emission. As a consequence of this feature, {\sl ROSAT} All-Sky Survey-based cluster samples are found to contain larger fractions of cool-core clusters than SZ-based samples, a phenomenon called ``cool core bias'' \citep[e.g.,][]{rossetti2017,Andrade-Santos2017}. The hypothesis that this is related to survey depth is supported by the recent results that the cool core fraction of eFEDS clusters is lower than that of the {\sl ROSAT} clusters, and is closer to SZ clusters \citep{2022Ghirardini}, probably because eFEDS has enough depth and better angular resolution to avoid the ``cool core bias'' found with {\sl ROSAT} clusters. These results indicate that the selection effects induced by the inconsistent survey depths of the two subsamples should be taken into account in our analysis.  

To construct two subsamples with consistent selection functions that are eligible for a direct comparison of concentration, we need to correct the above selection effects. 
This is performed with the following strategy. We only consider the clusters within the redshift range $0.05 < z < 0.6$ and the mass range $2\times10^{13}M_{\odot} < M_{500} < 8\times10^{14}M_{\odot}$ (see the left panel of Fig.~\ref{fig:sample}), where we have enough statistics for both supercluster members and isolated clusters. All the clusters in our sample have an exposure time $t$ within the range [60--2520~s], while the survey area rapidly decreases at deep exposures. With the cuts on mass and redshift, the sample size decreases from 8862 to 7658 (3577 SC members and 4081 isolated clusters). Then, we trim the two subsamples in the mass-redshift-exposure ($M-z-t$) space. 
The $M-z-t$ space is divided into $50\times50\times8=20000$ 3D grids. For mass and redshift, the boundaries of the grids are equally spaced in log- and linear-space, respectively. For exposure time, the following eight grids are manually chosen (in units of seconds): [60--150, 150--250, 250--350, 350--500, 500--700, 700--1000, 1000--1500, 1500--2520]. The edges are optimized to minimize each grid's width while including enough clusters. In each $M-z-t$ grid, we randomly select the same number of supercluster members and isolated clusters. In each randomization, 1302 unique clusters are selected for both supercluster members and isolated clusters. Since there are often different numbers of supercluster members and isolated clusters within one grid, the lists of clusters being selected in each randomization may differ from each other. To obtain stable results of selection, the randomization is repeated 1000 times, and the lists of clusters from each randomization are merged. We thus construct subsamples of supercluster members and isolated clusters, each consisting of $1302\times1000$ clusters. However, each subsample only contains about 1700 unique clusters, since most clusters are selected multiple times. Therefore, we weight each unique cluster in the two subsamples with the frequency of being selected when computing any average values. 

With the trimming approach described above, we establish two unbiased subsamples of clusters, namely, supercluster members and isolated clusters, with the same redshift distribution and mass distribution, and the same survey depth. Each subsample contains about 1700 clusters. Specifically, the median values of redshift, mass, and exposure time for SC members and isolated clusters for the trimmed subsamples are [0.2376, 0.2374], [1.90$\times 10^{14} M_{\odot}$, 1.89$\times 10^{14} M_{\odot}$], and [142.1s, 142.2s], respectively. Thus the differences are negligible and the two subsamples can be directly compared without significant selection effects. The consistency of the two subsamples is demonstrated in Fig.~\ref{fig:mzt_compare}. We do not apply the trimming process on other quantities except for redshift, mass, and exposure time. Other factors such as the discrepancy in Galactic hydrogen column density $n_{\rm H}$ of the two subsamples may also induce very minor impacts on the results. However, in the construction of the eRASS1 galaxy cluster sample, we have excluded most of the Galactic disk areas, and the majority of the clusters are located at high Galactic latitudes where the $n_{\rm H}$ is relatively low. Therefore, the effect of $n_{\rm H}$ should be negligible compared to the impacts of redshift, mass, and survey depth \citep{Kluge2024}.

\begin{figure}
\begin{center}
\includegraphics[width=0.49\textwidth, trim=5 15 65 65, clip]{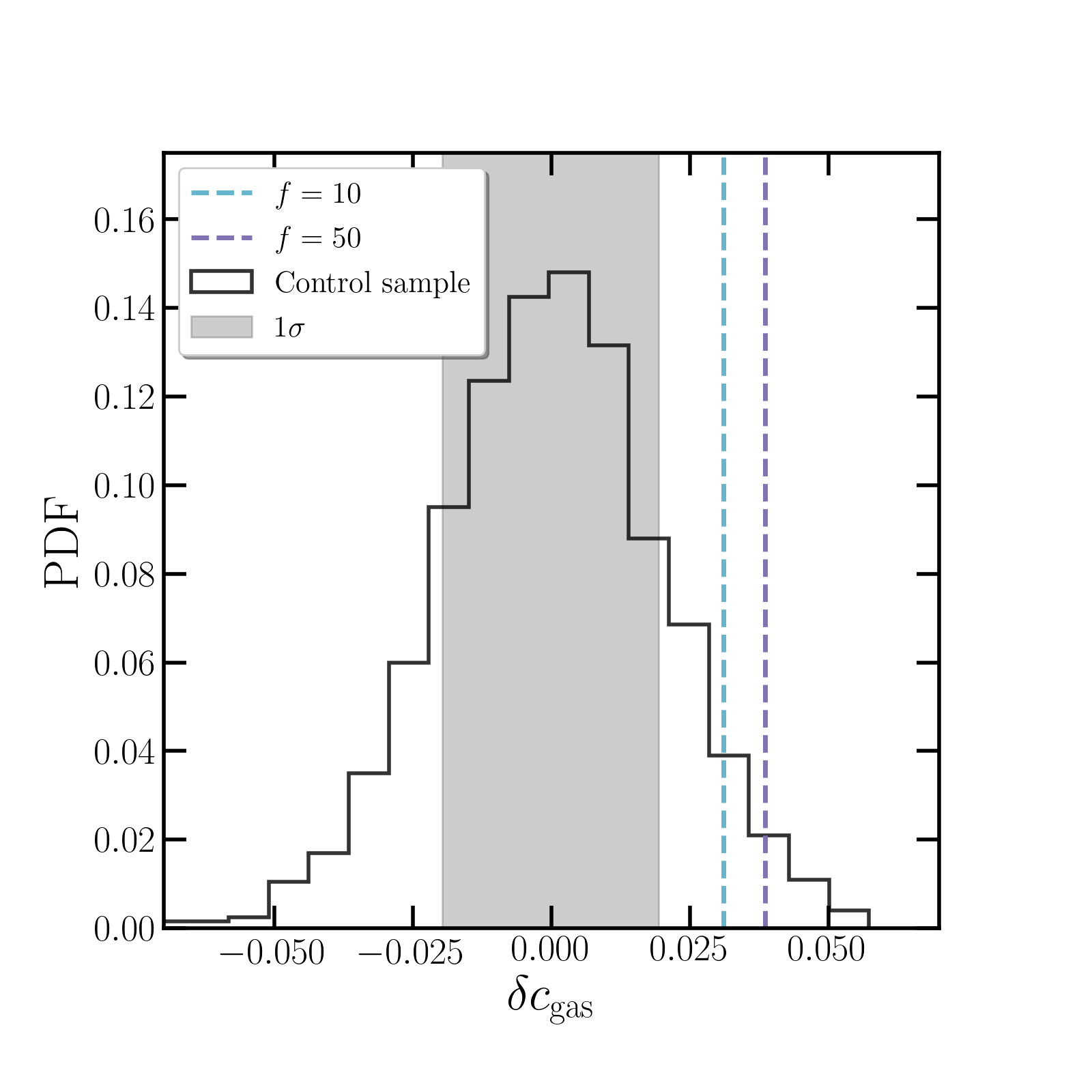}
\caption{\label{conf} Confidence level of the excess in gas mass concentration $c_{\rm gas}$. The histogram shows the distribution of $\delta c_{\rm gas}$ defined as $c_{\rm gas,S1}/c_{\rm gas,S2}-1$ when randomly dividing the clusters into two subsamples (S1 and S2) and repeating for 1000 times. The dashed lines indicate $\delta c_{\rm gas}$ when the clusters are divided into supercluster members and isolated clusters. The results are shown for different overdensity ratios in the identification of superclusters. The shaded area represents $1\sigma$ uncertainty. The cyan and purple lines correspond to 1.6$\sigma$ for $f=10$ and 2.0$\sigma$ for $f=50$, respectively. }
\end{center}
\end{figure}

\begin{figure}
\begin{center}
\includegraphics[width=0.49\textwidth, trim=5 15 65 65, clip]{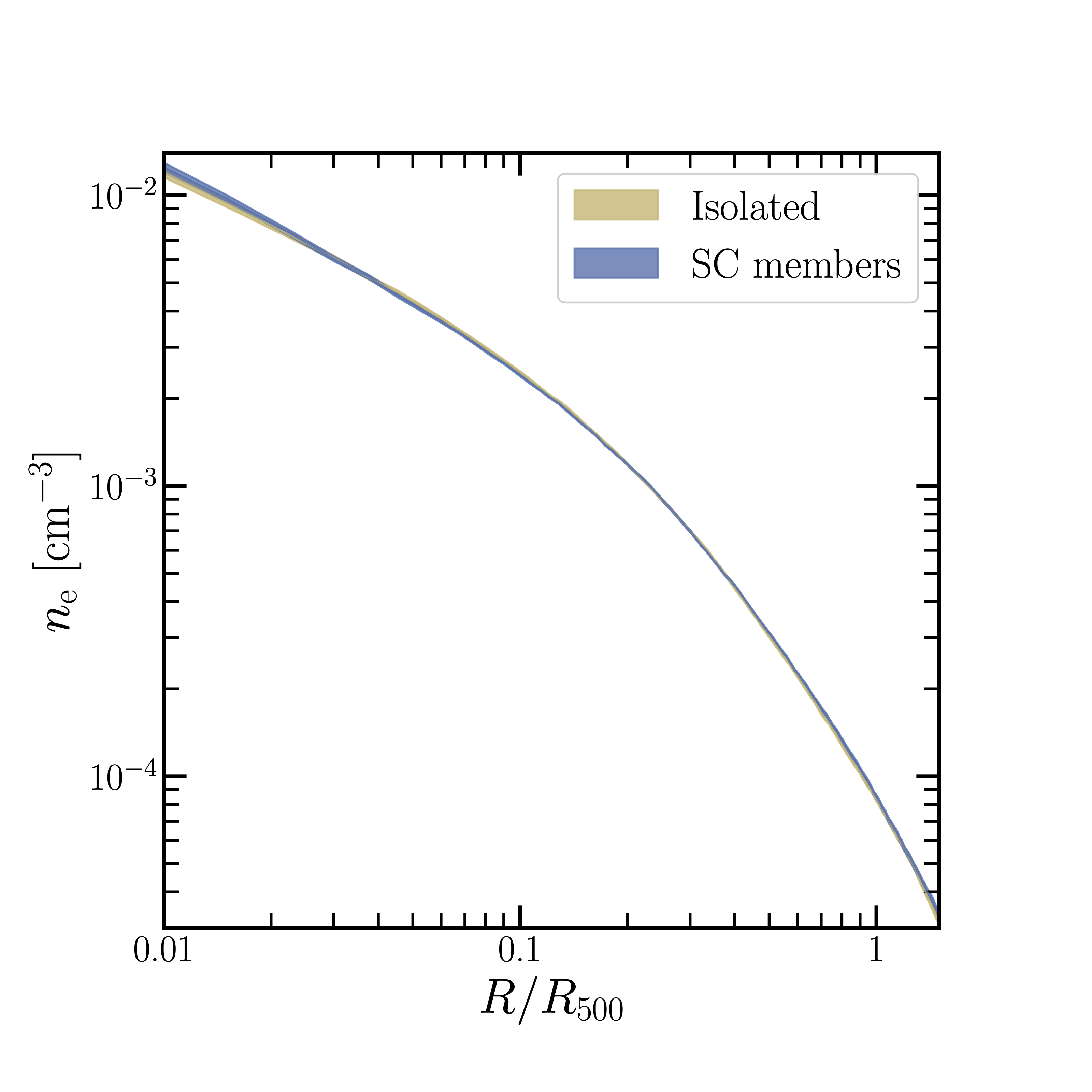}
\caption{\label{neprof50} Same as the left panel of Fig.~\ref{neprof} but for the supercluster sample of $f=50$.  }
\end{center}
\end{figure}

\section{Comparison of gas mass concentration}
\label{sec:results}
After obtaining the bias-free subsamples of supercluster members and isolated clusters, we can investigate their differences in halo properties. In this work, we adopt gas mass concentration as the proxy for halo assembly history. We simply define gas mass concentration as the ratio of gas mass within two radii: $c_{\rm gas} \equiv M_{\rm gas}(r_1)/M_{\rm gas}(r_2)$, where $r_1=0.3R_{500}$ and $r_2=R_{500}$. 

The $c_{\rm gas}$ for the clusters can be obtained from the X-ray analysis performed in \citet{Bulbul2024}. The eRASS1 X-ray data are processed with the {\sl eROSITA} Science Analysis Software System \citep[eSASS,][]{2022Brunner}\footnote{version {\tt eSASSusers\_211214\_0\_4}}. We use the tool MultiBand Projector in 2D \citep[MBProj2D,][]{2018Sanders}\footnote{https://github.com/jeremysanders/mbproj2d} to measure the gas properties of the eRASS1 clusters by fitting multi-band X-ray images. By forward-fitting background-included X-ray images of a galaxy cluster, MBProj2D provides the physical models of the cluster, such as the electron density profile, temperature profile, and metallicity profile \citep[see, e.g.,][for a recent application of MBProj2D]{Liu2023}. In the analysis of eRASS1 clusters, we divide the energy range [0.3--7]keV into seven bands (in units of keV): [0.3--0.6], [0.6--1.0], [1.0--1.6], [1.6--2.2], [2.2--3.5], [3.5--5.0], [5.0--7.0], to be able to constrain the ICM temperature. Counts images and exposure maps are created in these bands to be fitted with MBProj2D. When there are multiple clusters in the image, all the clusters are properly fitted instead of masked, to make sure that the background emission cannot be systematically boosted by the residual emission from neighboring clusters. This is particularly important for the analysis of supercluster members. We use all seven telescope modules (TMs) in the analysis. However, TMs 5 and 7 are ignored for the soft energy band below 1~keV, because they are affected by light leak \citep{2021Predehl}. The density profile is described using the model from \citet{vikhlinin2006a}, but without the second $\beta$ component:
\begin{equation}
n_{\mathrm p}n_{\mathrm e} = n_0^2 \cdot \frac{(r/r_c)^{-\alpha}}{(1+r^2/r_c^2)^{3\beta-\alpha/2}}\frac{1}{(1+r^{\gamma}/{r_s}^{\gamma})^{\epsilon/\gamma}}.
\end{equation}
$n_{\rm e}$ and $n_{\rm p}$ are the number densities of electron and proton, and we assume $n_{\rm e}=1.21n_{\rm p}$. $n_0$, $r_c$, $\alpha$, $\beta$, $\gamma$, $r_s$ and $\epsilon$ are free parameters. Gas density and gas mass profiles are then computed from the electron density profiles: $\rho_{\mathrm g}=n_{\rm e}m_{\rm p}A/Z$, where $A\sim1.4$ and $Z\sim1.2$ are the average nuclear charge and mass for ICM when assuming a metallicity of 0.3$Z_{\odot}$ \citep[e.g.,][]{mernier2018,2016Bulbul}. 

The median electron density profiles of supercluster members and isolated clusters are plotted in Fig.~\ref{neprof}. To quantify the difference between the two classes of clusters (if any), we construct two control samples by randomly dividing the clusters into two subsamples S1 and S2, and apply the same trimming process described above. This approach is repeated 1000 times, thus the standard deviation of the excess in $c_{\rm gas}$ of S1 over S2 represents the uncertainty in that of supercluster members and isolated clusters due to randomization. The median electron density profiles of S1 and S2 are also shown in Fig.~\ref{neprof}. 

\begin{figure*}
\begin{center}
\includegraphics[width=\textwidth, trim=0 0 0 58, clip]{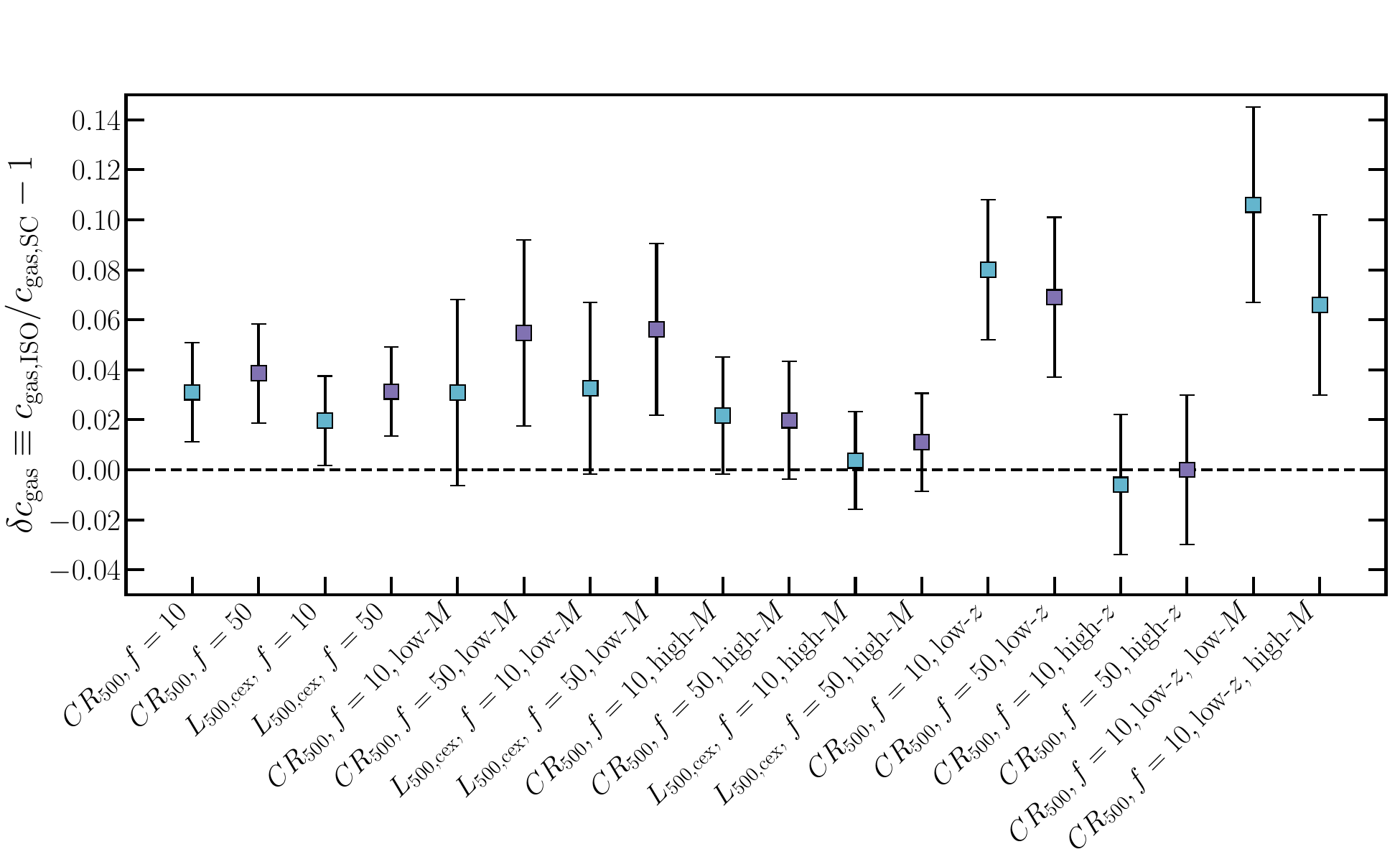}
\caption{\label{results} Summary of the results on $\delta c_{\rm gas}$ we obtained in this work for different choices of over density ratio $f$ in supercluster identification, cluster mass proxy, and mass range. The results for $f=10$ and $f=50$ are plotted in cyan and purple respectively, to show the increasing trend in $\delta c_{\rm gas}$ with $f$. The two measurements for total mass concentration obtained from weak-lensing analysis, $\delta c_{\rm tot}=0.149^{+0.109}_{-0.137}$ for $f=10$ and $\delta c_{\rm tot}=0.217^{+0.143}_{-0.149}$, are not shown in this plot because they are much larger than the results of gas mass concentration.  }
\end{center}
\end{figure*}

From Fig.~\ref{neprof}, we observe no obvious difference between supercluster members and isolated clusters in their electron density profiles. Quantitatively, the median $c_{\rm gas}$ is 0.160 and 0.155 for isolated clusters and supercluster members, respectively. Thus, if we quantify the HAB signal with the excess in $c_{\rm gas}$ of isolated clusters over supercluster members: $\delta c_{\rm gas} \equiv c_{\rm gas,ISO}/c_{\rm gas,SC}-1$, then we obtain $\delta c_{\rm gas}=0.031$. As a comparison, the standard deviation of $c_{\rm gas,S1}/c_{\rm gas,S2}-1$ is 0.019 (see Fig.~\ref{conf}). Therefore, our result on $\delta c_{\rm gas}$ is only at a confidence level of 1.6$\sigma$, corresponding to a marginal signal of HAB. 

We further investigate whether the low magnitude of HAB is due to the criterion we used to identify superclusters, namely, overdensity ratio $f=10$. In \citet{Liu2024}, we also provided a supplementary catalog of superclusters, identified with the same FoF method on the eRASS1 galaxy cluster catalog, but using a more strict criterion of overdensity: $f=50$. This supplementary catalog includes 929 supercluster systems and 2205 supercluster members. We perform the same selection and trimming approaches described in Sect.~\ref{sec:cluster} on the $f=50$ supercluster sample. After these processes, we obtain 956 supercluster members and 1269 isolated clusters for a direct comparison. We note that the notation ``isolated clusters'' here are the clusters that are not members of superclusters either in the $f=10$ catalog or the $f=50$ catalog. Namely, we exclude the clusters between $f=10$ and $f=50$, which are supposed to have intermediate clustering magnitudes. The median electron density profiles of the $f=50$ sample are shown in Fig.~\ref{neprof50}. Compared to the results of the $f=10$ sample, the enhancement in the difference of the electron density profiles is almost negligible. The median $c_{\rm gas}$ of isolated clusters and supercluster members are 0.159 and 0.153, respectively. The significance of $\delta c_{\rm gas}$ is increased from 1.6$\sigma$ to 2.0$\sigma$ (see Fig.~\ref{conf}). Therefore, we detect a subtle HAB signal with both the $f=10$ and $f=50$ supercluster samples.

It is also important to check if the signal is simply caused by merging clusters in the supercluster systems. We are more interested in the behavior of concentration as a proxy for cluster assembly history, instead of being a simple indicator of its current merging stage. Clusters that are currently undergoing merger activities naturally have more flattened gas distributions and are more likely to be identified as superclusters due to their small distances from neighbors. Therefore, we repeat our analysis by conservatively removing any cluster that intersects with a neighbor within their $R_{500}$ on the plane of the sky. Among the 8862 clusters in the sample, 741 clusters (about 8\%) are excluded. The minimum distance between any clusters in the remaining sample is larger than the sum of their $R_{500}$. Using this merger-free sample, we obtain $c_{\rm gas,ISO}=0.161$, $c_{\rm gas,SC}=0.156$, $\delta c_{\rm gas}=0.032\pm 0.021$ for $f=10$, and $c_{\rm gas,ISO}=0.161$, $c_{\rm gas,SC}=0.153$, $\delta c_{\rm gas}=0.051\pm 0.024$ for $f=50$. The results are in very good agreement with the full sample. We can thus rule out the hypothesis that the HAB signal is simply caused by clusters that are currently undergoing mergers.

\subsection{The impact of concentration on sample selection}
The masses of the eRASS1 clusters within $R_{500}$ are computed using the weak-lensing-calibrated scaling relation between the X-ray count rate within $R_{500}$ ($CR_{500}$), redshift, and mass $M_{500}$. This is very similar to the widely used $M-L_X$ scaling relation, but we use count rate instead of luminosity because the former is more correlated with the selection procedure of the eRASS1 clusters. Since the core emission is also included in the scaling relation, the mass computation can be affected by gas concentration.
In general, the masses of the clusters with higher gas mass concentrations can be slightly biased towards higher values, due to the X-ray surface brightness being proportional to $n_{\rm e}^2$ and thus is boosted by the bright core. This effect is minor and well below the typical errorbar of our mass measurements. However, since we aim to detect faint signals of only a few percent, we carefully check the impact of the mass-concentration correlation on our sample selection. The ideal solution for this issue is to measure the mass profile independently from the weak-lensing shear profile for each cluster in the sample. However, the signal-to-noise of individual shear profiles is too low for most of our clusters. Another commonly accepted solution is to exclude the core in the computation of scaling relations. This has been proved by recent studies, where the scatter in $M-L_X$ scaling relation is found to be smaller for core-excised luminosities \citep[see, e.g.,][]{bulbul2019}. 

Since our only aim is to avoid mass discrepancy between the two subsamples, there is no need to establish the $M-L_{\rm cex}$ scaling relation and re-compute the masses from core-excised luminosities. Given the two subsamples already have the same redshifts, the only necessary step is to make sure they are also consistent in core-excised luminosity $L_{\rm cex}$. Therefore, the trimming process on $L_{\rm cex}$ is already sufficient for our purpose.

We then repeat the selection and trimming processes described in Sect.~\ref{sec:cluster}, but in ($L_{\rm 500,cex}, z, t$) space instead of ($M-z-t$). In this work, we define the core size as $0.3R_{500}$ to be consistent with our definition of gas mass concentration. We divide the $L_{\rm 500,cex}$ range [$10^{41}-3\times 10^{44}$ erg/s] into 50 bins in log-space. The selections on redshift and exposure time remain the same as described in Sect.~\ref{sec:cluster}. The calculations based on control samples are also repeated accordingly. As a result, we obtain a median $c_{\rm gas}$ of 0.155 and 0.152 for isolated clusters and supercluster members, respectively. Therefore, the significance of concentration excess $\delta c_{\rm gas}=0.020\pm0.018$ is only at $1.1\sigma$. If we further investigate the $f=50$ supercluster sample, then we obtain $\delta c_{\rm gas}=0.031\pm0.018$. Therefore, the significance of $\delta c_{\rm gas}$ is increased to 1.7$\sigma$, close to the results obtained with count rate $CR_{500}$ as the mass proxy. These results also indicate that the mass-concentration correlation only has negligible impacts on our sample selection.

\subsection{Results for different mass and redshift ranges}

According to several previous works, HAB is stronger for low-mass halos than massive halos, or behaves differently in low- and high-mass regimes \citep[e.g.,][]{2006Wechsler}. Therefore, we further check the results when we divide the cluster sample into two mass ranges. We cut the sample at $M_{500}=2\times 10^{14} M_{\odot}$, and apply the same selection and trimming processes described in Sect.~\ref{sec:cluster}. We then perform the same experiments as for the whole sample. The $\delta c_{\rm gas}$ are obtained for different over density ratios $f=10$ and $f=50$, and for different mass proxies $CR_{500}$ and $L_{\rm 500,cex}$. Due to the cut on mass, the number of clusters is reduced by $\sim 50\%$ for each of the low-mass and high-mass samples compared to the whole sample. This further leads to a larger errorbar in $\delta c_{\rm gas}$. Therefore, it is expected that the significance of the results will be even lower than the whole sample, although the hints we obtain from the comparison between low- and high-mass regimes will still be useful.

In general, we find a decreasing trend in $\delta c_{\rm gas}$ with the investigated mass ranges. For the low-mass sample, we obtain larger $\delta c_{\rm gas}$ than the whole sample. For the high-mass sample, $\delta c_{\rm gas}$ is lower than both the low-mass sample and the whole sample, indicating that the difference in $c_{\rm gas}$ of supercluster members and isolated clusters is almost negligible for massive clusters above $2\times 10^{14} M_{\odot}$. However, this trend needs to be confirmed by further studies due to the large errorbar of the measurements (see Sect.~\ref{sec:discussions2}). 

We also investigate the results for different redshift ranges by dividing the whole sample into a low-$z$ sample ($0.05<z<0.2$) and a high-$z$ sample ($0.2<z<0.6$). $\delta c_{\rm gas}$ is enhanced to 6\%--8\% for the low-$z$ sample. Specifically, we obtain $\delta c_{\rm gas}=0.080\pm 0.028$ for $f=10$ and $\delta c_{\rm gas}=0.069\pm 0.032$ for $f=50$. As a comparison, the HAB signal completely disappears for the high-$z$ sample, where we obtain $\delta c_{\rm gas}=-0.006\pm 0.028$ for $f=10$ and $\delta c_{\rm gas}=0.000\pm 0.030$ for $f=50$. This redshift trend could partially originate from the mass trend because the low-$z$ sample contains most of the low-mass clusters, and the high-$z$ sample is dominated by massive clusters. However, another possibility is that the large-scale environment continuously affects the assembly of clusters, and the HAB signal tends to increase with time. We further split the low-$z$ clusters into a low-mass sample and a high-mass sample. Unsurprisingly, we obtain from the low-$z$ low-$M$ sample the largest $\delta c_{\rm gas}$ among all the experiments: $0.106\pm 0.039$. As a comparison, the low-$z$ high-$M$ sample has $\delta c_{\rm gas}=0.066\pm0.036$, lower than the result for low-$z$ low-$M$ clusters but still higher than that for the high-$M$ sample without redshift cut. These results, if tied together, suggest an increasing trend of $\delta c_{\rm gas}$ at both lower masses and lower redshifts. The only missing part in this scenario is the case of low-mass clusters at high redshifts, most of which are below the eRASS1 flux limit and not included in our sample. This will be improved in the future with deeper {\sl eROSITA} surveys (see Sect.~\ref{sec:discussions2}).

Our results of all the experiments described above are summarized in Fig.~\ref{results} and Table~\ref{tab:results}.

\begin{table*}
\caption{\label{tab:results} Summary of our results on $c_{\rm gas}$ and $c_{\rm tot}$ for supercluster members and isolated clusters. }
\begin{center}
\begin{tabular}[width=\textwidth]{cccccccccc}
\hline\hline
Mass range [$M_{\odot}$] & Redshift range & Mass proxy & $f$ & $N_{\rm SC}$ & $N_{\rm ISO}$ & $c_{\rm gas,SC}$ & $c_{\rm gas,ISO}$ & $\delta c_{\rm gas} $ & Confidence level \\
 & &  &  & & & $c_{\rm tot,SC}$ & $c_{\rm tot,ISO}$ & $\delta c_{\rm tot}$ &  \\
\hline
$2\times 10^{13}-8\times 10^{14}$ & 0.05--0.6 & $CR_{500}$ & 10 & 1687 & 1784 & 0.155 & 0.160 & 0.031$^{+0.019}_{-0.019}$ & 1.6$\sigma$ \\
$2\times 10^{13}-8\times 10^{14}$ & 0.05--0.6 & $CR_{500}$ & 50 & 956 & 1269 & 0.153 & 0.159 & 0.039$^{+0.020}_{-0.020}$ & 2.0$\sigma$ \\
$2\times 10^{13}-8\times 10^{14}$ & 0.05--0.6 & $L_{\rm 500,cex}$ & 10 & 1875 & 1951 & 0.152 & 0.155 & 0.020$^{+0.018}_{-0.018}$ & 1.1$\sigma$ \\
$2\times 10^{13}-8\times 10^{14}$ & 0.05--0.6 & $L_{\rm 500,cex}$ & 50 & 1058 & 1450 & 0.151 & 0.156 & 0.031$^{+0.018}_{-0.018}$ & 1.7$\sigma$ \\
$2\times 10^{13}-8\times 10^{14}$ & 0.05--0.6 & WL shear$^\dagger$ & 10 & 735 & 627 & 0.235 & 0.270 & 0.149$^{+0.109}_{-0.137}$ & 1.1$\sigma$ \\
$2\times 10^{13}-8\times 10^{14}$ & 0.05--0.6 & WL shear$^\dagger$ & 50 & 418 & 437 & 0.230 & 0.280 & 0.217$^{+0.143}_{-0.149}$ & 1.5$\sigma$ \\
$2\times 10^{13}-2\times 10^{14}$ & 0.05--0.6 & $CR_{500}$ & 10 & 928 & 899 & 0.160 & 0.165 & 0.031$^{+0.037}_{-0.037}$ & 0.8$\sigma$ \\
$2\times 10^{13}-2\times 10^{14}$ & 0.05--0.6 & $CR_{500}$ & 50 & 499 & 619 & 0.155 & 0.164 & 0.055$^{+0.037}_{-0.037}$ & 1.5$\sigma$ \\
$2\times 10^{13}-2\times 10^{14}$ & 0.05--0.6 & $L_{\rm 500,cex}$ & 10 & 836 & 747 & 0.152 & 0.156 & 0.033$^{+0.034}_{-0.034}$ & 1.0$\sigma$ \\
$2\times 10^{13}-2\times 10^{14}$ & 0.05--0.6 & $L_{\rm 500,cex}$ & 50 & 458 & 521 & 0.148 & 0.156 & 0.056$^{+0.034}_{-0.034}$ & 1.6$\sigma$ \\
$2\times 10^{14}-8\times 10^{14}$ & 0.05--0.6 & $CR_{500}$ & 10 & 748 & 869 & 0.152 & 0.155 & 0.022$^{+0.026}_{-0.026}$ & 0.9$\sigma$ \\
$2\times 10^{14}-8\times 10^{14}$ & 0.05--0.6 & $CR_{500}$ & 50 & 452 & 633 & 0.152 & 0.155 & 0.020$^{+0.024}_{-0.024}$ & 0.8$\sigma$ \\
$2\times 10^{14}-8\times 10^{14}$ & 0.05--0.6 & $L_{\rm 500,cex}$ & 10 & 794 & 869 & 0.149 & 0.150 & 0.004$^{+0.020}_{-0.020}$ & 0.2$\sigma$ \\
$2\times 10^{14}-8\times 10^{14}$ & 0.05--0.6 & $L_{\rm 500,cex}$ & 50 & 467 & 652 & 0.150 & 0.151 & 0.011$^{+0.020}_{-0.020}$ & 0.6$\sigma$ \\
$2\times 10^{14}-8\times 10^{14}$ & 0.05--0.2 & $CR_{500}$ & 10 & 658 & 595 & 0.145 & 0.156 & 0.080$^{+0.028}_{-0.028}$ & 2.8$\sigma$ \\
$2\times 10^{14}-8\times 10^{14}$ & 0.05--0.2 & $CR_{500}$ & 50 & 375 & 442 & 0.144 & 0.154 & 0.069$^{+0.032}_{-0.032}$ & 2.2$\sigma$ \\
$2\times 10^{14}-8\times 10^{14}$ & 0.2--0.6 & $CR_{500}$ & 10 & 1003 & 1170 & 0.162 & 0.161 & -0.006$^{+0.028}_{-0.028}$ & -0.2$\sigma$ \\
$2\times 10^{14}-8\times 10^{14}$ & 0.2--0.6 & $CR_{500}$ & 50 & 563 & 803 & 0.162 & 0.162 & 0.000$^{+0.030}_{-0.030}$ & 0.0$\sigma$ \\
$2\times 10^{13}-2\times 10^{14}$ & 0.05--0.2 & $CR_{500}$ & 10 & 463 & 404 & 0.144 & 0.159 & 0.106$^{+0.039}_{-0.039}$ & 2.7$\sigma$ \\
$2\times 10^{14}-8\times 10^{14}$ & 0.05--0.2 & $CR_{500}$ & 10 & 192 & 187 & 0.146 & 0.156 & 0.066$^{+0.036}_{-0.036}$ & 1.8$\sigma$ \\
\hline
\end{tabular}
\tablefoot{$\dagger:$ The results obtained from weak lensing shear profiles are the concentrations of total mass instead of gas mass. }
\end{center}
\end{table*}

\section{Comparison of total mass concentration}
\label{sec:wl}
To compare with the gas mass concentration results, we also attempt to use total mass concentration as the proxy of halo assembly history. Due to the different dynamic characteristics of the collisionless dark matter particles and the viscous intracluster gas, the absolute values of the total mass and gas mass concentrations are not comparable with each other, but the trends in $\delta c_{\rm gas}$ and $\delta c_{\rm tot}$ are expected to be consistent. Although the weak-lensing shear profiles for individual clusters are too noisy to measure concentration, we can use the stacked shear profile to obtain the average mass concentration for a sample of clusters. We perform this experiment for the two samples with $f=10$ and $f=50$, after applying the same trimming process described in Sect.~\ref{sec:cluster}. We use the public Dark Energy Survey year 3 (DES-Y3) data to measure the stacked shear profiles for supercluster members and isolated clusters, respectively. The numbers of SC members and isolated clusters in the DES footprint are 735 and 627 for $f=10$, 418 and 437 for $f=50$.

In the weak-lensing formalism, the tangential gravitational shear, $\gamma_{\rm t}$, induced by the foreground structure (lens; galaxy clusters in this case) is related to the surface density of the lens, $\Sigma(R)$, as
\begin{equation}
    \Delta\Sigma (R) = \gamma_{\rm t}(R) \Sigma_{\rm crit} (z_{\rm l}, z_{\rm s}),
\end{equation}
where
\begin{equation}
    \Sigma_{\rm crit} = \frac{c^2}{4 \pi G} \frac{D_{\rm A,s}}{D_{\rm A,l} D_{\rm A, ls}}
\end{equation}
and
\begin{equation}
    \Delta\Sigma (R) = \Bar{\Sigma}(<R) - \Sigma(R).
\end{equation}
Here, $z_{\rm l}$ and $z_{\rm s}$ are the redshifts of the lens and the source. $D_{\rm A,l}$, $D_{\rm A,s}$ and $D_{\rm A,ls}$ are the angular diameter distances to the lens, to the source, between the lens and the source, respectively. $\Bar{\Sigma}(<R)$ is the mean surface density within the radius $R$ from the center of the lens. Note that the tangential shear is related to the two shear components, $\gamma_1$ and $\gamma_2$, as
\begin{equation}
    \gamma_{\rm t} = -\gamma_1 \cos(2\theta) -\gamma_2 \sin(2\theta),
\end{equation}
where $\theta$ is the position angle of the galaxy with respect to the x-axis of the system.

For the weak-lensing measurement, we use the shape catalog from the DES-Y3 data measured by \textsc{Metacalibration} algorithm \citep{Sheldon2017,Huff2017,Gatti2021}. In \textsc{Metacalibration}, the true shear, $\boldsymbol{\gamma}$, is related to the measured galaxy shape, $\boldsymbol{e}$ through the response matrix, $\boldsymbol{\mathcal{R}}$, 
\begin{equation}
    \langle \boldsymbol{\gamma} \rangle = \langle \boldsymbol{\mathcal{R}}\rangle^{-1} \langle \boldsymbol{e} \rangle.
\end{equation}
We also calculate and include the selection response, $\boldsymbol{\mathcal{R}_{\rm s}}$, which arises due to the specific selection of galaxies. We refer the readers to the above studies for further details. 

We adopt the photometric redshifts of the galaxies measured by the \textsc{DNF} algorithm taken from the DES-Y3 Gold catalog \citep{Sevilla-Noarbe2021}. We make use of the two different redshift values for each galaxy as given by the catalog -- the mean estimated redshift ($z_{\rm mean}$) and a redshift value randomly chosen from the probability distribution ($z_{\rm MC}$). For further details, we refer the readers to \citep{Sevilla-Noarbe2021} and the references therein.

In principle, the measured gravitational shear by \textsc{Metacalibration} could be multiplicatively biased. Furthermore, due to the large uncertainty of the photometric redshift of the source galaxies, the weak-lensing measurement is subject to another multiplicative bias through $\Sigma_{\rm crit}$ \citep[e.g.][]{McClintock2019}. However, since our main goal is to compare the concentration (not the mass) of the supercluster members to that of the isolated clusters, and those multiplicative biases apply to all samples equally, we do not include these bias factors in our analysis. We also ignore the uncertainty on the cluster redshift since it is significantly smaller than that of the source galaxies.

Following \citet{McClintock2019} and \citet{Shin2021}, we construct our estimator for $\Delta\Sigma (R)$ as
\begin{equation}
\label{eq:DeltaSigmaEstimator}
    \Delta\tilde{\Sigma} (R) = \frac{\sum_{ij} w^{ij} e_{\rm t}^{j} (R) }{\sum_{ij} w^{ij} \Sigma^{-1}_{\rm c,MC}(z^i_{\rm l},z^j_{\rm s}) (\mathcal{R}_{\rm t}^j+\mathcal{R}_{\rm s})},
\end{equation}
where
\begin{equation}
    w^{ij} = \frac{1}{(\sigma_{\gamma}^j)^2} \Sigma^{-1}_{\rm c,mean}(z^i_{\rm l},z^j_{\rm s}).
\end{equation}
Here, $i$ runs over the lens clusters, and $j$ runs over the source galaxies. $e_{\rm t}$ is the tangential component of the measured galaxy shape. Also, $\Sigma^{-1}_{\rm c,MC}(z_{\rm l},z_{\rm s})$ is evaluated with the redshift of the source galaxies using $z_{\rm MC}$, while for $\Sigma^{-1}_{\rm c,mean}(z^i_{\rm l},z^j_{\rm s})$ $z_{\rm mean}$ is used. In addition, $\mathcal{R}_{\rm t}$ is the tangential component of the response matrix rotated to the tangential coordinate, and $\sigma_{\gamma}$ is the uncertainty of the shape measurement. We refer the readers for this estimator's details and validation to \citet{McClintock2019}. 

To minimize the contamination from the foreground or cluster member galaxies that do not carry any lensing signal, we select source galaxies such that $z_{\rm s} > z_{\rm l} + 0.1$ using the photo-z from \textsc{DNF}. However, due to the high uncertainties on the photometric redshifts of the galaxies, the cluster member galaxies that do not carry any lensing signal could leak into our source galaxy selection, which dilutes the weak-lensing signal. Therefore, one must measure and take into account this scale-dependent systematic bias in the analysis. This is generally called the ``boost factor'', $\mathcal{B}(R)$, which we model following the method outlined in \citet{Gruen2014} and \citet{Varga2019} as follows.

For a measured photometric redshift distribution at the cluster-centric distance of $R$, $P(z|R)$, we decompose the distribution into the cluster contamination, $P_{\rm cont}(z)$, and the true background distribution, $P_{\rm bg}(z)$:
\begin{equation}
    P(z|R) = f_{\rm cl}(R) P_{\rm cont}(z) + (1-f_{\rm cl}(R)) P_{\rm bg}(z).
\end{equation}
Note that $f_{\rm cl}(R)$, a free parameter per radial bin, represents the fraction of member galaxy contamination in the distribution, and we model $P_{\rm cont}(z)$ as a radius-independent Gaussian distribution with a mean ($\mu_{\rm z,cl}$) and a standard deviation ($\sigma_{\rm z,cl}$) as additional free parameters.
The corrected weak-lensing profile after accounting for the boost factor becomes,
\begin{equation}
    \Delta\Bar{\Sigma}(R)_{\rm corr} = \mathcal{B}(R) \Delta\Bar{\Sigma}(R) = \frac{1}{1-f_{\rm mis}}\Delta\Bar{\Sigma}(R).
\end{equation}

We model the 3D cluster mass profile as,
\begin{equation}
    \rho(r) = \rho_{\rm NFW}(r) + b_{\rm c} \rho_{\rm m}(z) \xi_{\rm mm}(r|z),
\end{equation}
where $\rho_{\rm NFW}=\rho_{\rm crit}\delta_{\rm c}/[(r/r_{\rm s})(1+r/r_{\rm s})^2]$ is a Navarro-Frenk-White profile \citep{1997Navarro} for which we have two free parameters: the total mass, $M_{500}$, and the concentration, $C_{500}=R_{500}/r_{\rm s}$. The second term is so called the 2-halo contribution due to the nearby halos that are clustered, where $\rho_{\rm m}(z)$ is the mean matter density of the Universe at the redshift $z$, $\xi_{\rm mm}(r|z)$ is the matter-matter correlation function of the Universe at $z$ with the separation of $r$, and $b_{\rm c}$ is the large-scale halo bias \citep{Tinker2010} that is another free parameter in our model. We then integrate this 3D density model into 2D along the line of sight to obtain the 2D projected surface mass density, $\Sigma(R)$.

On the other hand, the X-ray centers used here could be offset from the true centers of the halos. Therefore, we include the miscentering effect in our model. 
For a given set of clusters, the stacked surface density profile in the presence of miscentering could be expressed as,
\begin{equation}
    \Sigma(R) = (1-f_{\rm mis}) \Sigma_0(R) + f_{\rm mis} \Sigma_{\rm mis}(R),
\end{equation}
where $f_{\rm mis}$ is the fraction of the miscentered clusters, $\Sigma_0$ the surface density profile without any miscentering, and $\Sigma_{\rm mis)}$ the profile with miscentering, which could be expressed as,
\begin{equation}
    \Sigma_{\rm mis} (R) = \int dR_{\rm mis} P(R_{\rm mis}) \Sigma_{\rm mis} (R|R_{\rm mis}).
\end{equation}
here, $R_{\rm mis}$ is the miscentered distance, and we assume a Rayleigh distribution for $P(R_{\rm mis})$, 
\begin{equation}
    P(R_{\rm mis}) = \frac{R_{\rm mis}}{\sigma^2_{\rm R}} \exp{-\frac{R^2_{\rm mis}}{2\sigma^2_{\rm R}}}. 
\end{equation}
By geometry, one can also show that the profile of a halo that is miscentered by $R_{\rm mis}$ is, 
\begin{equation}
    \Sigma_{\rm mis}(R|R_{\rm mis}) = \int_0^{2\pi} \frac{d\theta}{2\pi} \Sigma_0\Big(\sqrt{R^2+R^2_{\rm mis}+2RR_{\rm mis}\cos(\theta)}\Big).
\end{equation}
\citet{Grandis2024} provides the average miscentering properties for the {\sl eROSITA} clusters. However, since our goal is to compare the properties of supercluster members and isolated clusters, for each of which the miscentering properties are not well studied, we adopt flat priors for the two miscentering parameters, $f_{\rm mis}=[0,1]$ and $\ln \sigma_{\rm R}=[0.05,0.5]$.

\begin{figure*}
\begin{center}
\includegraphics[width=0.495\textwidth]{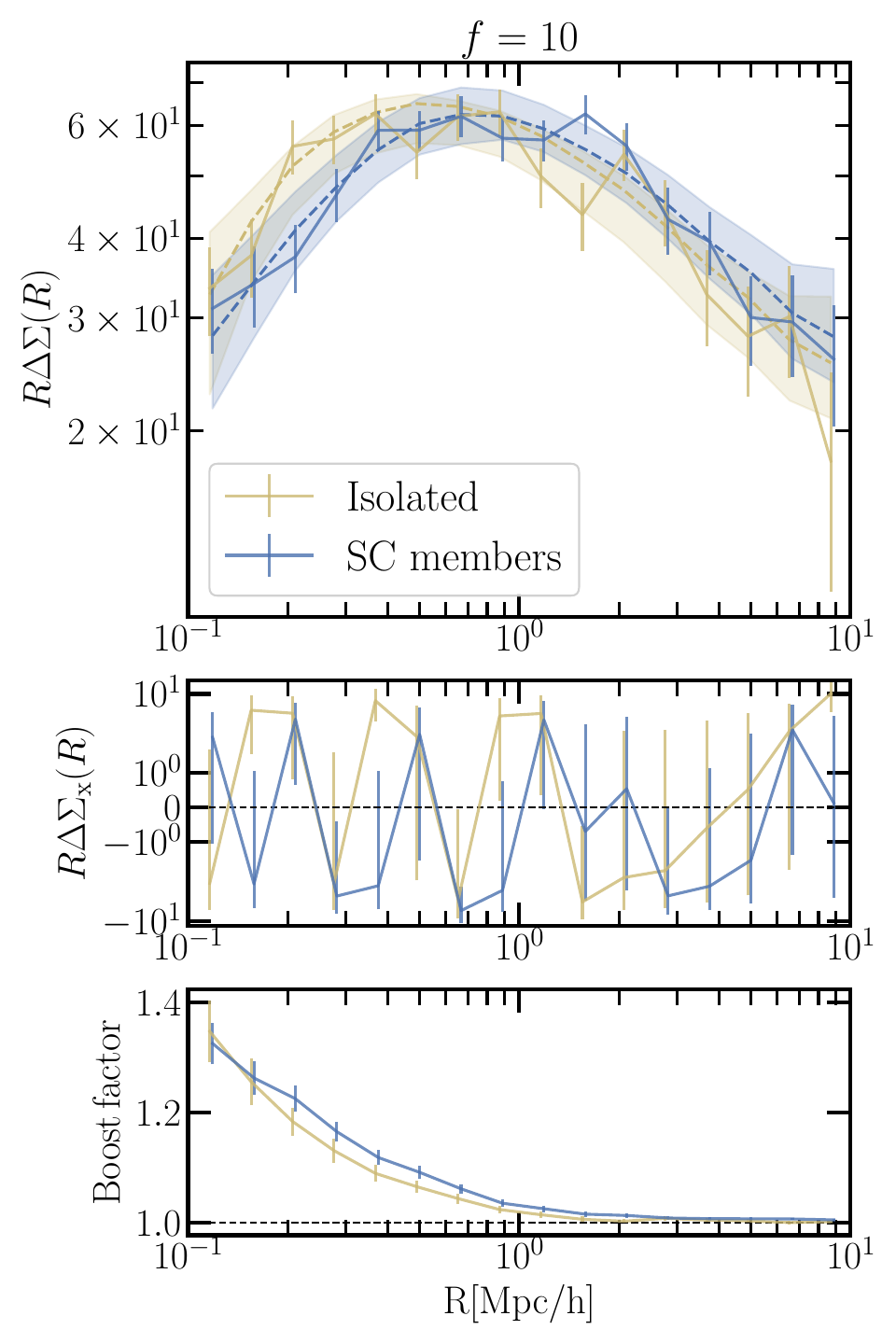}
\includegraphics[width=0.495\textwidth]{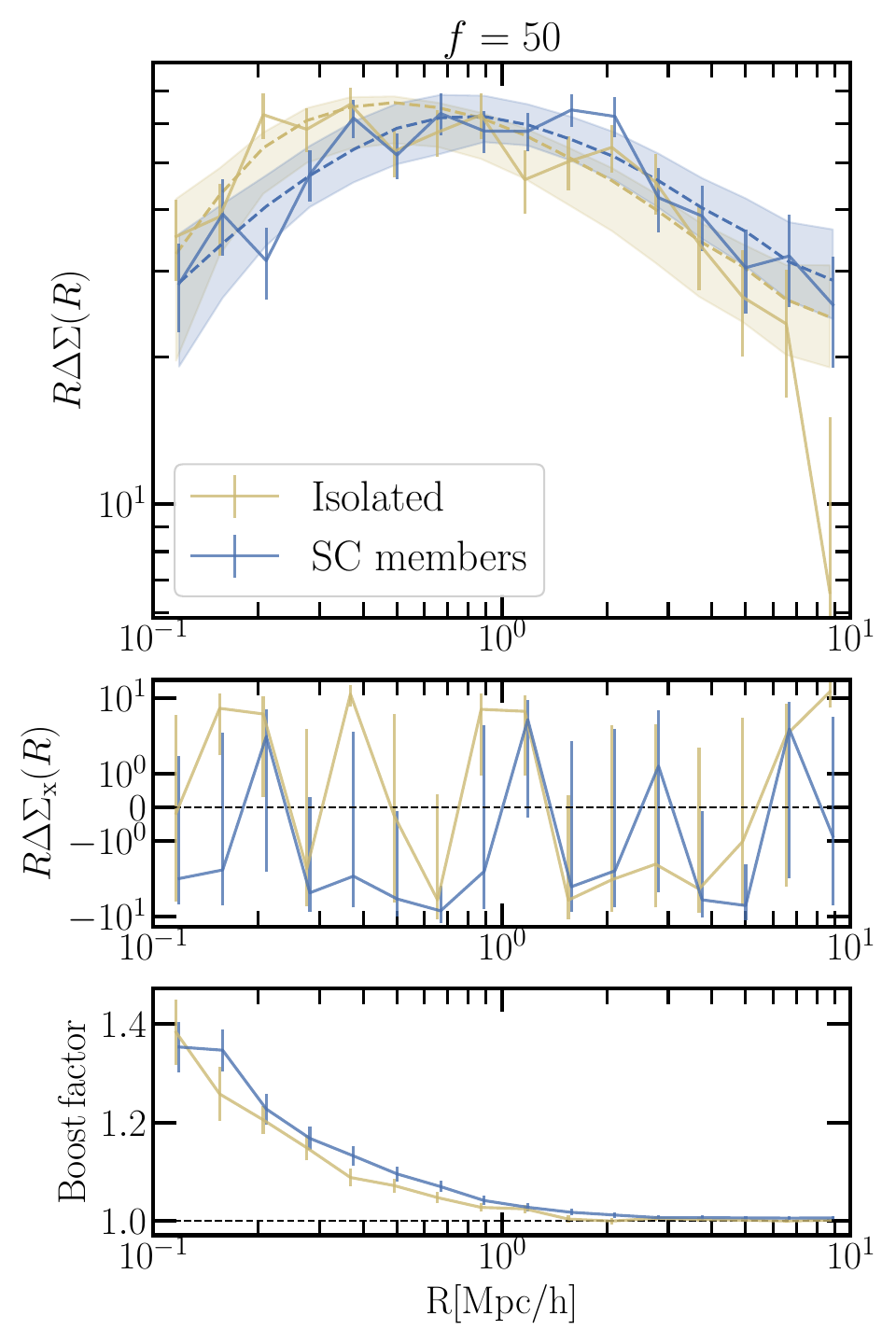}
\caption{\label{fig:wl-measurement} The measured weak-lensing $\Delta\Sigma$ profiles (top), the cross components for a null test (middle), and the boost factors (bottom) for the $f=10$ sample (left) and the $f=50$ sample (right). The supercluster members are plotted in blue and the isolated clusters in yellow. The shaded regions represent the 68\% confidence best-fit interval, while the dashed lines correspond to the best-fit models.}
\end{center}
\end{figure*}

We measure $\Delta\tilde{\Sigma}(R)$ (Eq.~\ref{eq:DeltaSigmaEstimator}) in 16 logarithmically spaced radial bins between 0.1 and 10.0 $h^{-1}{\rm Mpc}$ in physical unit. The $P(z|R)$ of each radial bin for the boost factor fitting is calculated with redshift bins between $z$=0.05 and 1.25 with the bin width of $\Delta z$=0.05. $P_{\rm bg}(R)$ is calculated around the random positions within the survey footprint ($N$=30 times the number of clusters).

The covariance matrix of $\Delta\tilde{\Sigma}(R)$ is obtained by bootstrapping the clusters with $100,000$ bootstrapped samples. For each bootstrapped cluster sample, we obtain the best-fit boost factor parameters ({$\mu_{\rm z,cl}$, $\sigma_{\rm z,cl}$, $f_{\rm cl}(R_i)$}) by fitting the observed $P(z|R)$ to the model described above. Then this bootstrapped boost factor is multiplied to the corresponding bootstrapped $\Delta\tilde{\Sigma}(R)$, from which we obtain our final boost-factor-corrected $\Delta\tilde{\Sigma}_{\rm corr}(R)$ with its covariance matrix. 

We then perform MCMC analyses on the corrected weak-lensing profile to get the constraints on the halo properties, for which we have five free parameters, $M_{500}$, $C_{500}$, $b_{\rm c}$, $f_{\rm mis}$ and $\sigma_{\rm R}$. We fix the redshift of the model to the mean cluster redshift and assume a Gaussian likelihood for the fitting. We use \textsc{emcee} package \citep{MCMC} for obtaining the MCMC chains.

The measured weak-lensing profiles with the boost factor correction as a function of cluster-centric radius are shown in Fig.~\ref{fig:wl-measurement}. The top panels show the $\Delta\Sigma$ profiles for the $f=10$ sample (left) and the $f=50$ sample (right), in both of which the isolated clusters (yellow) show higher amplitudes below $0.7 h^{-1}{\rm Mpc}$ than that of the supercluster members (blue), hinting at a higher value of concentration for the isolated clusters. The 68\% confidence intervals of the best-fit model for each measurement are shown with the shaded bands with the corresponding colors, with the best-fit models plotted with the dashed lines of the same colors. For the $f=10$ ($f=50$) sample, the best-fit $\chi^2$ per degree of freedom for the supercluster members is 9.0/11 (14.9/11) corresponding to the probability to exceed (PTE) of 0.62 (0.18), while that for the isolated clusters is 14.7/11 (16.7/11) corresponding to PTE=0.20 (0.12). In the middle panels, we show as a null test the cross-components of the weak-lensing measurements, defined as,
\begin{equation}
    \gamma_{\rm \times} = -\gamma_1 \sin(2\theta) +\gamma_2 \cos(2\theta),
\end{equation}
substituting $\gamma_{\rm t}$, which must be consistent with zero for the isotropic lens. The minimum of $\chi^2$ per degree of freedom for $f=10$ ($f=50$) sample is $12.1/16$ ($12.9/16$) for the supercluster members and $16.1/16$ ($21.3/16$) for the isolated clusters, which are consistent with zero. Finally, the bottom panels show the measured boost factors with the same color scheme, where the supercluster members and the isolated clusters show features distinct from each other, while they both asymptote to 1 at large scales as expected. We defer a detailed analysis of the difference in the galaxy number density profile between the two samples, as suggested by the boost factors, to future studies.

In Appendix~\ref{app:corner}, we show the corner plots of the MCMC chains run for the $f=10$ (Fig.~\ref{fig:corner_f10}) and $f=50$ samples (Fig.~\ref{fig:corner_f50}). One could notice that for both cases the supercluster members and the isolated clusters exhibit similar masses ($M_{500}$), agreeing well within 1-$\sigma$ with each other, while the isolated clusters have larger concentration ($C_{500}$) than that of the supercluster members, in alignment with our previous findings. Using these chains, we then obtain the concentration of the total mass $c_{\rm tot}$ with the same definition of $c_{\rm gas}$ ($c_{\rm tot}=M(<0.3R_{500})/M_{500}$). For the $f=10$ supercluster sample, the total mass concentrations $c_{\rm tot}$ of the supercluster members and isolated clusters are $c_{\rm tot,SC}=0.235^{+0.023}_{-0.013}$ and $c_{\rm tot,ISO}=0.270^{+0.021}_{-0.014}$, respectively, corresponding to $\delta c_{\rm tot} = 0.149^{+0.109}_{-0.137}$. For the $f=50$ supercluster sample, we obtain $c_{\rm tot,SC}=0.230^{+0.023}_{-0.017}$, $c_{\rm tot,ISO}=0.280^{+0.027}_{-0.017}$, with $\delta c_{\rm tot}$ of $0.217^{+0.143}_{-0.149}$. As expected, the absolute values of total mass concentrations are larger than gas mass concentrations, since dark matter particles are collisionless and intracluster medium being viscous. On the other hand, for both cases, we observe the same trend as gas mass concentration, that supercluster members are also less concentrated than isolated clusters in total mass. The significance level of $\delta c_{\rm tot}$ is 1.5$\sigma$, slightly lower than that of $\delta c_{\rm gas}$, possibly because only $<$40\% of the clusters are within the DES footprint and included in the weak lensing analysis. We also note that the posteriors for the miscentering parameters are wide and poorly constrained, so that by marginalizing over those parameters we enhance the robustness of our finding at the cost of less precision. It is also noticeable that the posteriors for the large-scale halo bias parameter ($b_{\rm c}$) are almost identical between supercluster members and isolated clusters, indicating that the HAB is not detected at the large-scale halo bias level, but only at the concentration level. We note, however, that our analysis is limited to $R<10h^{-1}$Mpc, and therefore may not be suitable for probing such large-scale bias signals. We defer a more detailed study on this matter to a future study.

\section{Discussions}
\label{sec:discussions}
In this section, we provide possible interpretations of our findings and discuss potential improvements to our results in the near future when deeper eRASS data are available.

\subsection{Interpretation of the results}

As presented in the previous sections, in all the experiments we have performed, we consistently observe a trend that supercluster members have lower concentrations, both in gas mass and total mass, than isolated clusters. This result is rather stable over a series of choices on cluster mass proxy, supercluster overdensity ratio, and cluster mass range. We also find that the concentration excess of isolated clusters compared to supercluster members ($\delta c_{\rm gas}$ and $\delta c_{\rm tot}$) increases marginally when we choose a higher threshold on the overdensity ratio of supercluster identification, implying that the difference in concentration is likely related to environmental effect. 
Although the significance level of any single experiment is not high enough (the maximum significance of $\delta c_{\rm gas}$ is $2.8\sigma$), these results are indeed in line with our current understanding of HAB based on many previous works on numerical simulations, where HAB is detected for massive halos. In these works, halos with higher clustering magnitudes are found to be less concentrated \citep[e.g.,][]{2006Wechsler,2007Wetzel,2007Gao}. 

To understand and interpret the signal we detected in this work, we first revisit the physical origin of HAB proposed by previous works. As already mentioned in Sect.~\ref{sec:intro}, a widely accepted explanation for HAB of low-mass halos is that the tidal fields of the neighboring massive halos prevent the low-mass halos from accreting matters. For example, \citet{2007Wang} found that low-mass old halos are often located close to more massive halos. The gravitational tidal fields of the neighboring massive halo produce a hot environment \citep[e.g.,][]{2005Mo}, which suppresses the mass accretion of the low-mass halos, and leads to the old ages of these low-mass halos. In this scenario, the environmental effect should generally decrease with halo mass. More massive halos dominate their vicinity and their evolution is less affected by the environment. This implies that the signal of HAB should be much weaker in clusters, which is consistent with our results. Moreover, the fact that we are observing systematically larger $\delta c_{\rm gas}$ in the low-mass sample, despite the large errorbar, is also in line with the above scenario.

Another scenario is that supercluster members are less concentrated simply because they are located in denser regions and thus have experienced more mergers during their assembly history. Merging processes between clusters can significantly flatten the distributions of both the baryon and dark matter components, thus supercluster members are expected to have lower concentrations than isolated clusters in both the gas mass and total mass. Compared to low-mass clusters, massive clusters are less affected by minor mergers, while the occurrence of major mergers is much rarer compared to minor mergers, which provides an explanation of the trend that $\delta c_{\rm gas}$ marginally decreases with mass. Moreover, since mergers between clusters occur continuously, it can be inferred that the strength of HAB signal increases with time, which is consistent with the larger $c_{\rm gas}$ we find for low-$z$ clusters. Therefore, the environmental dependence of cluster merger rates can also interpret our results and should be regarded as another possible origin of HAB. 

\subsection{Expected improvements with deeper eRASS data and multi-wavelength cluster surveys}
\label{sec:discussions2}
The next question is how can we improve the significance of the detection of HAB. An effective way to increase the significance of the signal is to expand the sample of supercluster members. {\sl eROSITA} has completed more than four all-sky surveys. About $5\times 10^4$ clusters and groups are expected to be detected in eRASS:4. Using the same criteria in supercluster identification, we expect to expand the sample of supercluster members by a factor of four. With the deeper X-ray data, the uncertainty in the measurements of electron density profiles will also be reduced significantly. Therefore, with the eRASS:4 data and cluster sample, we expect to improve the significance of the HAB signal by at least a factor of three, corresponding to a detection at $6\sigma$ confidence level for the full sample without mass and redshift cuts. eRASS:4 will also significantly expand the sample of low-mass clusters at high redshifts, and further confirm the mass and redshift dependencies of HAB we find in this work.

\begin{figure}
\begin{center}
\includegraphics[width=0.49\textwidth, trim=5 15 65 65, clip]{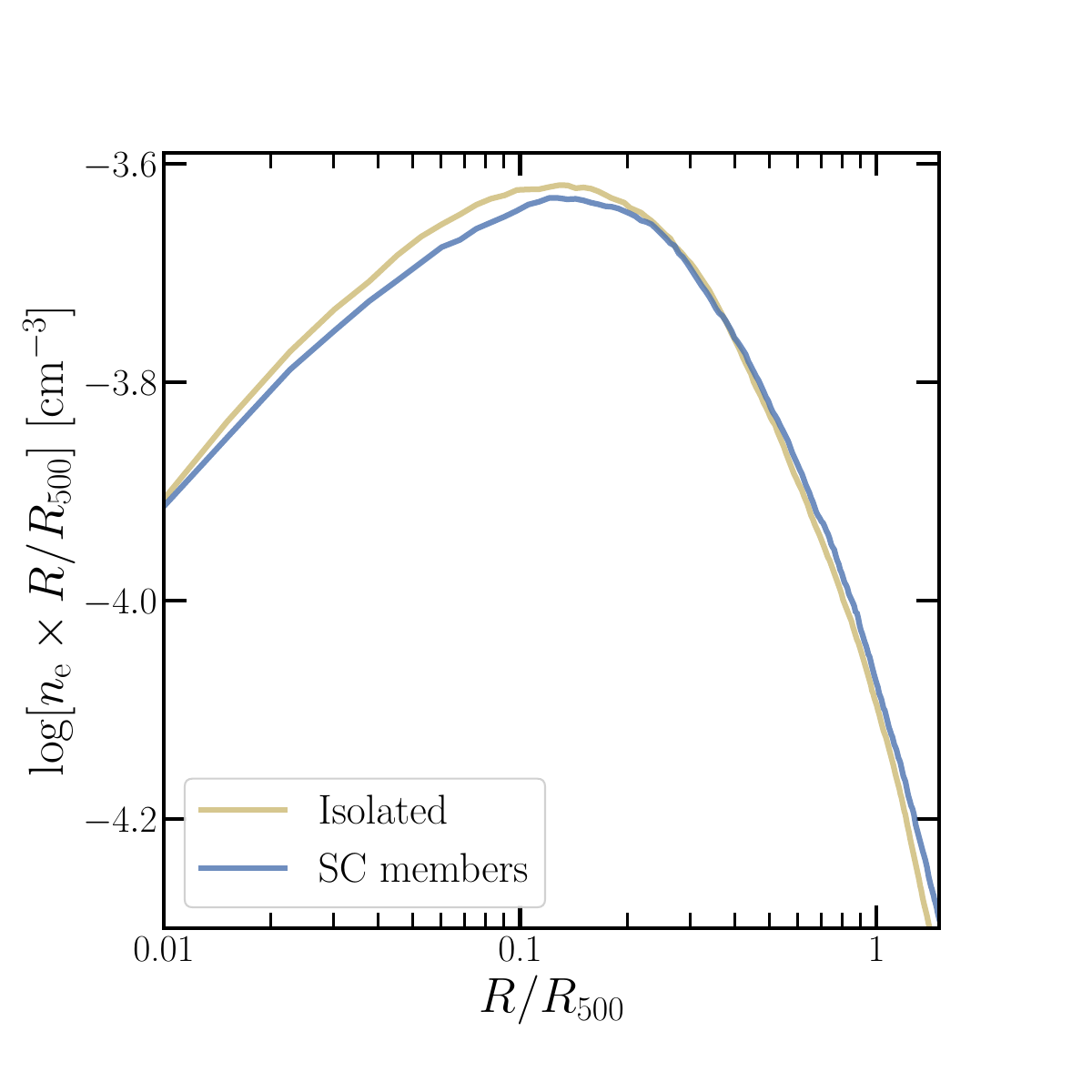}
\caption{\label{fig:example} Same as the left panel of Fig.~\ref{neprof} but without considering the difference in exposure time of the supercluster members and isolated clusters. The y-axis is re-scaled by the radius to reduce the dynamic range of the plot and improve the visibility of the difference between the two samples. The median electron density profile of supercluster members (blue) shows an apparently lower concentration than that of the isolated clusters (yellow). However, this trend should not be simply interpreted as the signal of halo assembly bias. It is mostly caused by the fact that supercluster members are located in areas with higher survey depth, where low-concentration clusters have a higher probability of being detected. This plot demonstrates the importance of correcting selection effects in the analysis of HAB. }
\end{center}
\end{figure}

Another approach is to make joint analyses by including the samples of clusters and superclusters detected in other wavelengths, such as optical and mm/submm (Sunyaev–Zeldovich effect, SZ). Cluster samples selected in different bands have their advantages and disadvantages. Optical cluster samples are more complete than X-ray and SZ samples, although the purity is usually lower, and the selection function is more complicated. SZ cluster samples contain fewer contaminants (e.g., projected interlopers in optical and AGN in X-ray), and have a simpler selection function directly correlated with cluster mass. However, limited by the much lower angular resolution of SZ telescopes, SZ observations provide fewer constraints compared to X-rays on cluster properties such as gas mass concentration. Therefore, combining these cluster samples and the multi-wavelength observations would help establish a more comprehensive dataset to study HAB with superclusters. We note that an essential step in such an approach is to carefully account for the selection effects in each band and each survey. As an example, we show in Fig.~\ref{fig:example} how sensitive is the detection of HAB to the selection effects in our eRASS1 X-ray survey. In this figure, we plot the electron density profiles of the supercluster members and the isolated clusters obtained in the same way as plotted in the left panel of Fig.~\ref{neprof}, but without applying the trimming process on exposure time. Namely, the two subsamples are consistent in both redshift and mass but have different survey depths. In this case, the difference in gas mass concentration is visible in the plot and would be misinterpreted as the ``signal'' of HAB if we do not exclude the selection effect caused by the inhomogeneity in survey depth. This plot showcases the importance of correcting selection effects in the exploration of HAB with large-area surveys.

\section{Conclusions}
\label{sec:conclusions}
We propose a novel method to explore halo assembly bias in cluster-sized halos. The essence of this method is using superclusters to identify the clusters that are located in denser environments and with higher clustering magnitudes (namely, supercluster members) and compare their assembly history with that of isolated clusters, to investigate the environmental effect on halo assembly history. In this work, we apply this method to the largest-ever X-ray galaxy cluster and supercluster samples obtained from the first {\sl eROSITA} all-sky survey \citep{Bulbul2024,Kluge2024,Liu2024}. Using the eRASS1 supercluster catalog, we construct two subsamples of galaxy clusters which consist of supercluster members and isolated clusters respectively. A ``trimming'' approach is employed on the two subsamples to overcome selection effects and ensure the two subsamples are consistent in redshift, mass, and survey depth. The concentration of the intracluster medium, defined as the ratio of gas mass within 0.3$R_{500}$ and $R_{500}$, is adopted as the proxy of halo assembly history. The signal of halo assembly bias is quantified with the excess in gas mass concentration of supercluster members compared to supercluster members: $\delta c_{\rm gas} \equiv c_{\rm gas,ISO}/c_{\rm gas,SC}-1$. We obtain $\delta c_{\rm gas}$ under multiple conditions, including different choices on the overdensity ratio for supercluster identification, cluster mass proxies, and cluster mass ranges. We also perform weak lensing analysis on the two subsamples. By stacking the weak lensing shear profiles of the clusters in each subsample, we compare the concentrations of the total mass, $\delta c_{\rm tot}$, defined in the same way as gas mass concentration.

We find that the average gas mass concentration of isolated clusters is a few percent larger than that of supercluster members. The result of $\delta c_{\rm gas}$ varies with the choices on supercluster overdensity ratio $f$ and mass ranges, but remains positive in all the cases. Specifically, we measure slightly larger $\delta c_{\rm gas}$ when adopting the supercluster sample identified with higher $f$, which, by definition, corresponds to higher clustering magnitudes and selects clusters in even denser environments. These results are also supported by the comparison of total mass concentrations between the two samples, measured from the weak lensing analysis, which shows similar trends obtained from gas mass. By dividing our cluster sample into subsamples based on mass and redshift, we also find that $\delta c_{\rm gas}$ is larger for clusters with lower mass and at lower redshifts. Our findings are consistent with the prediction of halo assembly bias in cluster-scale, where halos located in denser environments are less concentrated than isolated ones, and this trend is stronger for less massive and low-redshift halos. These phenomena can be interpreted by the fact that clusters in denser environments such as superclusters have experienced more mergers than isolated clusters in their assembling history.

Among all the measurements we obtained in this work, the maximum confidence level of $\delta c_{\rm gas}$ or $\delta c_{\rm tot}$ is $2.8\sigma$, only corresponding to a marginal signal of HAB. Therefore, as the first attempt to explore HAB with superclusters, our results are not conclusive enough to end the debate on the existence of HAB for cluster-sized halos. In the near future, the results of this work are expected to be improved with deeper {\sl eROSITA} surveys which will significantly expand the sample of both supercluster members and isolated clusters, and improve the quality of the X-ray data. This work paves the way to explore HAB with superclusters, and demonstrates that large samples of superclusters, a category of object in the Universe that has been known for decades, can advance our understanding of the evolution of large-scale structure. With the methodology we have laid out in this work, a joint analysis combining optical and SZ cluster and supercluster samples will also provide useful constraints on HAB in cluster scale. However, we stress that the selection effects of different cluster surveys must be carefully accounted for in the investigation of HAB.

\begin{acknowledgement}
We greatly thank the anonymous referee for his/her constructive comments. AL thanks Daisuke Nagai, Stefano Borgani, and Stefano Ettori for helpful discussions. This work is based on data from eROSITA, the soft X-ray instrument aboard SRG, a joint Russian-German science mission supported by the Russian Space Agency (Roskosmos), in the interests of the Russian Academy of Sciences represented by its Space Research Institute (IKI), and the Deutsches Zentrum für Luft- und Raumfahrt (DLR). The SRG spacecraft was built by Lavochkin Association (NPOL) and its subcontractors and is operated by NPOL with support from the Max Planck Institute for Extraterrestrial Physics (MPE).

The development and construction of the eROSITA X-ray instrument was led by MPE, with contributions from the Dr. Karl Remeis Observatory Bamberg \& ECAP (FAU Erlangen-Nuernberg), the University of Hamburg Observatory, the Leibniz Institute for Astrophysics Potsdam (AIP), and the Institute for Astronomy and Astrophysics of the University of Tübingen, with the support of DLR and the Max Planck Society. The Argelander Institute for Astronomy of the University of Bonn and the Ludwig Maximilians Universität Munich also participated in the science preparation for eROSITA.
\\
The eROSITA data shown here were processed using the eSASS/NRTA software system developed by the German eROSITA consortium.
\\
Ang Liu, Esra Bulbul, Vittorio Ghirardini, and Xiaoyuan Zhang acknowledge financial support from the European Research Council (ERC) Consolidator Grant under the European Union’s Horizon 2020 research and innovation program (grant agreement CoG DarkQuest No 101002585).
\\
Nicola Malavasi acknowledges funding by the European Union through a Marie Sk{\l}odowska-Curie Action Postdoctoral Fellowship (Grant Agreement: 101061448, project: MEMORY). Views and opinions expressed are however those of the author only and do not necessarily reflect those of the European Union or of the Research Executive Agency. Neither the European Union nor the granting authority can be held responsible for them.
\\
This work made use of SciPy \citep{jones_scipy_2001}, Matplotlib, a Python library for publication-quality graphics \citep{Hunter2007}, Astropy, a community-developed core Python package for Astronomy \citep{Astropy2013}, NumPy \citep{van2011numpy}. 

\end{acknowledgement}

\bibliography{aliu}

\begin{appendix}
\section{Constraints of the halo parameters from the weak-lensing analysis}
\label{app:corner}

\begin{figure*}
\begin{center}
\includegraphics[width=0.99\textwidth]{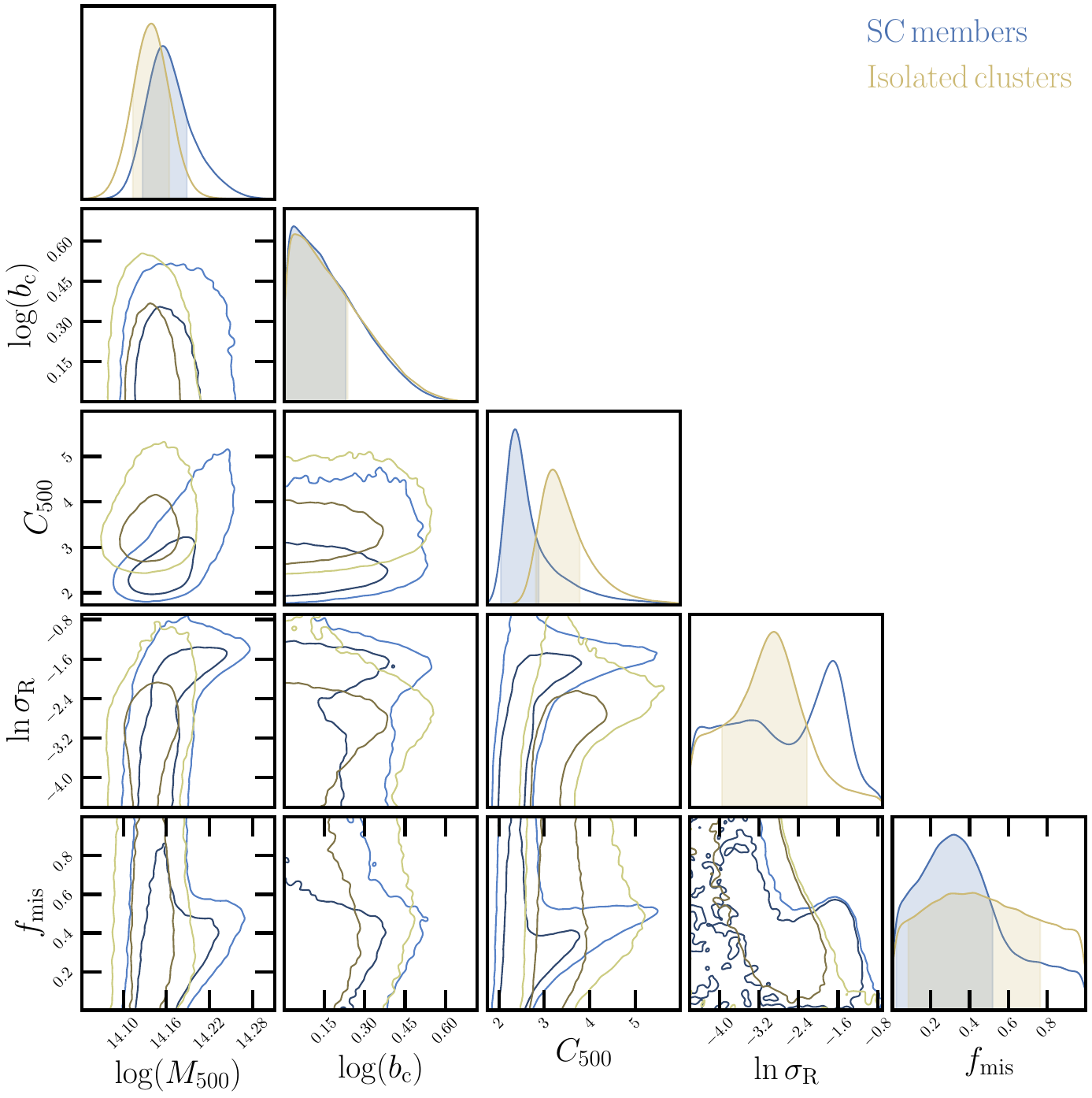}
\caption{\label{fig:corner_f10}The 1- and 2-$\sigma$ contours and the marginalized distributions of the halo parameters from the MCMC fitting (Sec.~\ref{sec:wl}) for the $f=10$ samples (blue: supercluster members, yellow: isolated clusters).}
\end{center}
\end{figure*}

\begin{figure*}
\begin{center}
\includegraphics[width=0.99\textwidth]{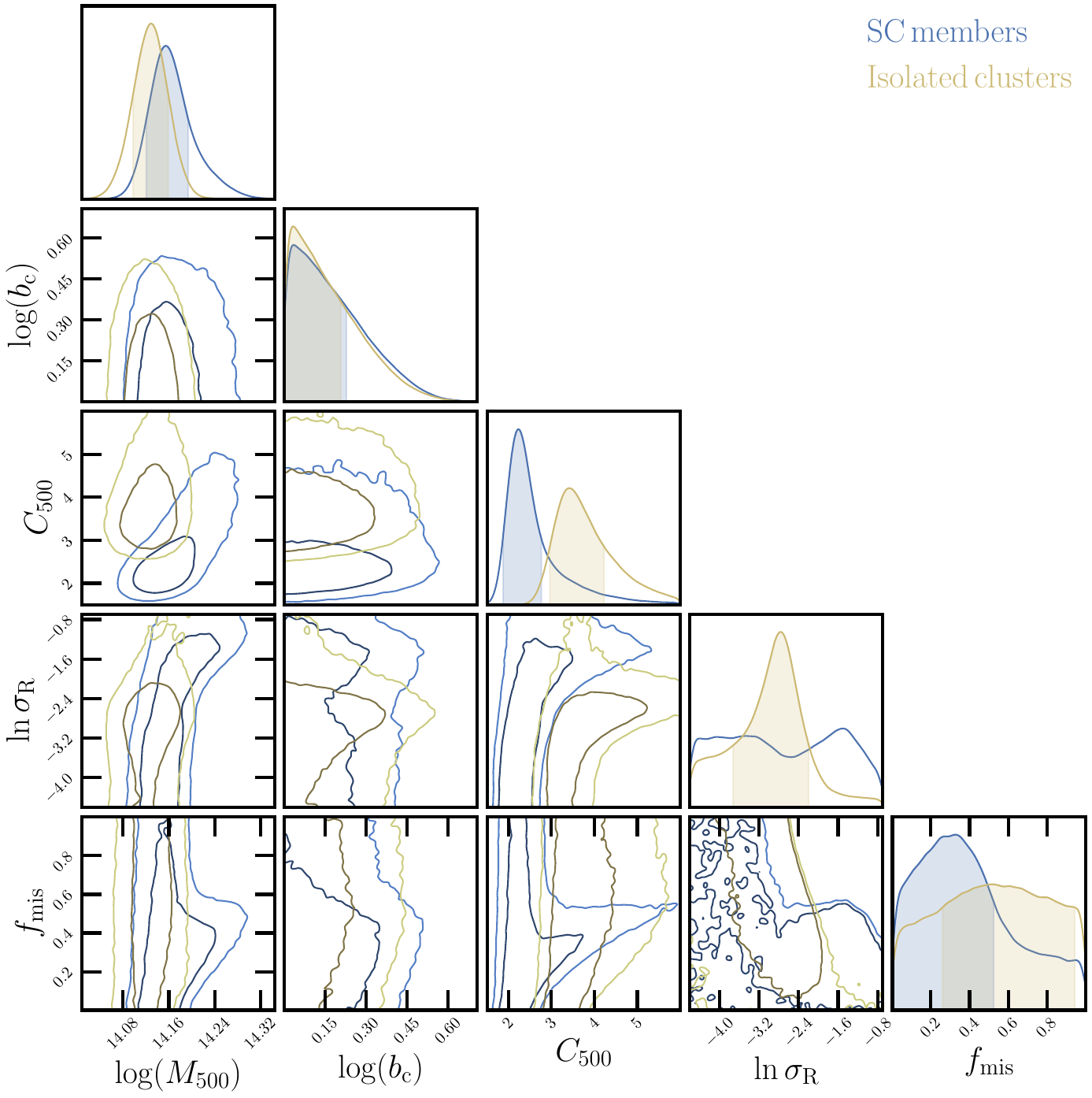}
\caption{\label{fig:corner_f50}Similar to Fig.~\ref{fig:corner_f10}, for the $f=50$ sample.}
\end{center}
\end{figure*}

In Fig.~\ref{fig:corner_f10} and Fig.~\ref{fig:corner_f50}, we show the constraints from the MCMC fitting of the weak-lensing profiles around our cluster samples (see Sect.~\ref{sec:wl} for details).

\end{appendix}

\end{document}